\documentstyle[12pt]{article}
\def\bit {\begin{itemize}}
\def\eit {\end{itemize}}
\def\beq {\begin{equation}}
\def\eeq {\end{equation}}
\def\bef {\begin{figure}}
\def\ef {\end{figure}}

\def\ben {\begin{enumerate}}
\def\een {\end{enumerate}}
\def\bec {\begin{center}}
\def\ec {\end{center}}
\def\bet {\begin{tabbing}}
\def\et {\end{tabbing}}
\def\beqa {\begin{eqnarray}}
\def\eeqa {\end{eqnarray}}
\def \omd {i\omega_c\delta/\tau}

\def \rprm {^{y_r^+}_{y_r^-}}
\begin{document}
\topmargin=-1cm

\begin{center}

{\bf \Large Fast global oscillations in
networks of integrate-and-fire neurons with low firing rates}

\vspace{0.5cm}

{\bf Nicolas Brunel\footnote{email: brunel@lps.ens.fr} and Vincent 
Hakim\footnote{email: hakim@lps.ens.fr}}

\vspace{0.5cm}

LPS\footnote{Laboratory associated with CNRS, Paris 6 and Paris 7
Universities}, Ecole Normale Sup{\'e}rieure

24 rue Lhomond, 75231 Paris Cedex 05, France

\end{center}

\begin{abstract}
We study analytically the dynamics of a network of sparsely connected
inhibitory integrate-and-fire neurons in a regime where individual
neurons emit spikes irregularly and at a low rate. In the limit when
the number of neurons $N\rightarrow\infty$, the network exhibits a
sharp transition between a stationary and an oscillatory global
activity regime where neurons are weakly synchronized. The activity
becomes oscillatory when the inhibitory feedback is strong enough.
The period of the global oscillation is found to be mainly controlled
by synaptic times, but depends also on the characteristics of the
external input.  In large but finite networks, the analysis shows that
global oscillations of finite coherence time generically exist both
above and below the critical inhibition threshold.  Their
characteristics are determined as functions of systems parameters, in
these two different regimes.  The results are found to be in good
agreement with numerical simulations.
\end{abstract}
\today

\section{Introduction}

Oscillations are ubiquitous in neural systems and have been the focus
of several recent studies (for reviews see e.g.~Gray 1994, Singer and
Gray 1995, Buzs{\'a}ki and Chrobak 1995, Ritz and Sejnowski 1997).  In
particular, fast global oscillations in the gamma frequency range ($>
30$ Hz) have been reported in the visual cortex (Gray et al 1989, Eckhorn
et al 1993, Kreiter and Singer 1996), in the olfactory cortex (Laurent and
Davidowitz 1994) and in the hippocampus (Bragin et al 1995).  Even
faster oscillations (200Hz) occur in the hippocampus of the rat
(Buzs{\'a}ki et al 1992, Ylinen et al 1995).  In some experimental data,
(see e.g.~Eckhorn et al 1993, Csicsvari et al 1998, Fisahn et al 1998)
individual neuron recordings show irregular spike emission, at a rate
which is low compared to the global oscillation frequency\footnote{
Fast oscillations may be due in some cases to a synchronized subset of
cells with high firing rates. The observation of cells with the
required property has been recently reported in (Gray and McCormick
1996).}.  This raises the question of whether a network composed
of neurons firing irregularly at low rates can exhibit fast collective
oscillations, which theoretical analyses and modelling studies may
help to answer.

Previous studies of networks of spiking neurons have mostly analyzed,
or simulated, synchronized oscillations in regimes in which neurons
behave themselves as oscillators, with interspike intervals strongly
peaked around their average value (see e.g.~Mirollo and Strogatz 1990,
Abbott and van Vreeswijk 1993, van Vreeswijk et al 1994, Gerstner
1995, Hansel et al 1995, Gerstner et al 1996, Wang and Buzs{\'a}ki 1996,
Traub et al 1996). Several oscillatory regimes have been found with
either full or partial synchronization.  A regime particular to
globally coupled systems has been described where the network breaks
into a few fully synchronized clusters (Golomb and Rinzel 1994, van
Vreeswijk 1996).  In some simulations of networks with detailed
biophysical characteristics, cells fire sparsely and irregularly
during a global oscillation (Traub et al 1989, Kopell and LeMasson
1994, Wang et al 1995), but the complexity of individual neurons in
these models makes it difficult to clearly understand of the origin of
the phenomenon.  The possible appearance of fast oscillations in a
network where all neurons fire irregularly with an average frequency
which is much lower than the population frequency therefore remains an
intriguing question.  It is the focus of the present work.

Recurrent inhibition plays an important role in the generation of
synchronized oscillations as shown by in vivo (McLeod and Laurent
1996) and in vitro experiments (Whittington et al 1995) in different
systems.  This has been confirmed by several modelling studies (van
Vreeswijk et al 1994, Gerstner et al 1996, Wang and Buzs{\'a}ki 1996,
Traub et al 1996).  It has also been recently shown using simple
models that networks in which inhibition balance excitation (Tsodyks
and Sejnowski 1995, Amit and Brunel 1997a, van Vreeswijk and
Sompolinsky 1996) are naturally composed of neurons with low and
irregular firing.  Simulations (Amit and Brunel 1997b) have shown
that, in one such model composed of sparsely connected
integrate-and-fire (IF) neurons, the highly irregular single neuron
activity is accompanied by damped fast oscillations of the global
activity.

In order to study the coexistence of individual neurons with low
firing rates and fast collective oscillations in its simplest setting,
we analyze in the present paper a sparsely connected network entirely
composed of identical inhibitory IF neurons.  Our aim is to provide a
clear understanding of this type of synchrony and to precisely
determine :\\ - i) under which conditions collective excitations of
high frequencies arise in such networks\\ - ii) what controls the
different characteristics (amplitude, frequency, coherence time,...)
of the global oscillation.

Simulation results are presented first which shows that the essence of
the phenomenon is present even in this simple system. Both the neurons
firing rates and the auto-correlation of the global activity are very
similar to those reported in (Amit and Brunel 1997b). 

We begin by presenting simple arguments which give an estimation of
the firing rate of individual neurons and the frequency of the global
oscillation and which lead to think that the global oscillation only
appears above a well-defined parameter threshold.

In order to make the analysis more precise and complete,
we then generalize the analytic approach of Amit and Brunel (1997a)
which was restricted to the computation of firing rates in stationary
states.  The sparse random network connectivity leads
the firing patterns of different neurons to be only weakly
correlated. 
As a consequence, the network state can be described by the
instantaneous distribution of membrane potentials of the neuronal
population, together with the firing probability in this
population. 
We obtain the coupled temporal evolution equations for
these quantities, the time-independent solution of which coincides
with the stationary solution of (Amit and Brunel 1997a).

A linear stability analysis shows that this time-independent solution
becomes unstable only when the strength of recurrent inhibition exceeds a
critical level, in agreement with our simple arguments.
When this critical level is reached, the stationary
solution becomes unstable and an oscillatory solution develops (via a
Hopf bifurcation).  The time scale of the period of the corresponding
global oscillations is set by a synaptic time,
independently of the firing rate of individual neurons, but the period
precise value also depends on the characteristics of the external input.

The analysis is then pushed to higher orders. We obtain a reduced
evolution equation describing the network collective dynamics.  The
effects coming from the finite size of the network are also discussed.
We show that having a large but finite number of neurons gives a small
stochastic component to the collective evolution equation. As a
result, it is shown that cross-correlations in a finite network
present damped oscillations both above and below the critical
inhibition level.  Below the critical level, the noise controls the
oscillation amplitude which decreases as the number of neurons is
increased (at a fixed number of connections per neuron).  Above the
critical level, the main effect of the noise is to produce a phase
diffusion of the global oscillation. An increase in the number of
neurons results in an increase of the global oscillation coherence
time and in a reduced damping in average cross-correlations.

Finally, the effect of some of our simplifying assumptions is studied.
We shortly discuss the effect of allowing variability in synaptic
times and number of synaptic connections from neuron to neuron. We
also consider the effect of introducing a more detailed description of
postsynaptic currents into the model. The technical aspects of our
computations are detailed in several appendices.

\section{Description of the network and simulations}

\begin{figure}
\setlength{\unitlength}{1cm}
\begin{picture}(14,7)
\put(-2,-8){\includegraphics{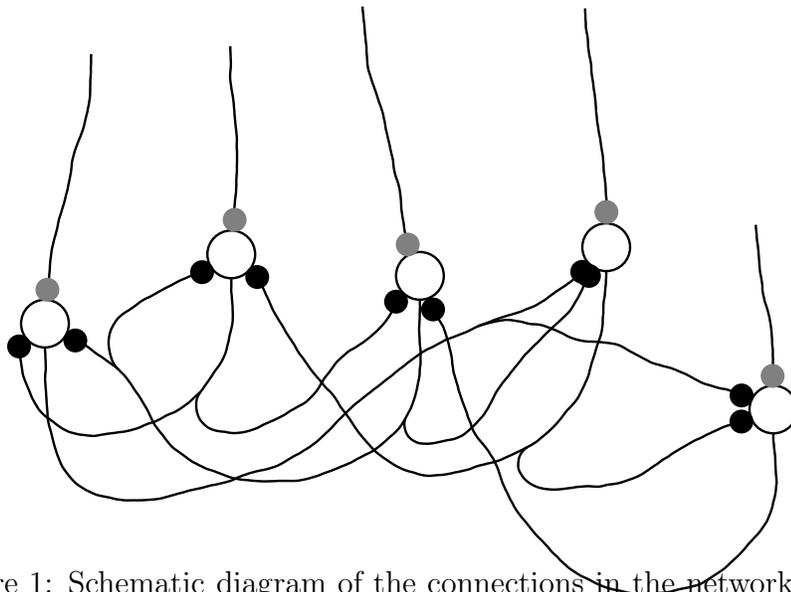}}
\end{picture}
\caption{Schematic diagram of the connections in the network of $N$
neurons; each neuron (indicated as an open disk) receives $C$
inhibitory connections (indicated as black) from within the network
and $C_{ext}$ excitatory connections (indicated as grey) from neurons
outside the network.}
\label{networkmap}
\end{figure}

We analyse the dynamics of a network composed of $N$ identical
inhibitory single compartment integrate-and-fire (IF) neurons.  Each
neuron receives $C$ randomly chosen connections from other neurons in
the network.  It also receives $C_{ext}$ connections from excitatory
neurons outside the network (see Fig.~\ref{networkmap}). We consider a
sparsely connected case with $\epsilon=C/N \ll 1$.

\begin{figure}
\setlength{\unitlength}{1cm}
\begin{picture}(14,6)
\put(0,-2){\includegraphics{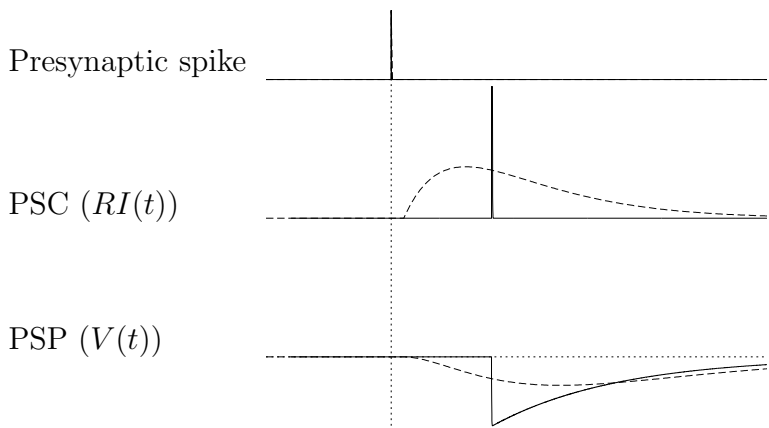}}
\put(0,5){Presynaptic spike}
\put(0,3.1){PSC ($RI(t)$)}
\put(0,1.3){PSP ($V(t)$)}
\end{picture}
\caption{Comparison of the synaptic response characteristics in our
model and in a more realistic model.  The top trace shows the
presynaptic spike. The middle trace shows the corresponding
postsynaptic current (PSC).  The bottom trace shows the corresponding
postsynaptic potential (PSP) for a neuron initially at resting
potential.  Full lines: our model, in which the synaptic current is
described by a delta function a time $\delta$ after the presynaptic
spike. Dashed lines: a more realistic synaptic response, in which the
PSC is described by an $\alpha$-function with latency (transmission
delay) $\tau_L$ and synaptic time constant $\tau_S$
$(t-\tau_L)\exp(-(t-\tau_L)/\tau_S)/\tau_S$.  Our synaptic
characteristic time $\delta$ can roughly be identified with the sum of
latency and synaptic decay time, $\tau_L+\tau_S$.  See the discussion
in Section \ref{section:synaptic}.}
\label{IPSP}
\end{figure}

Each neuron is simply described by its membrane potential. Let us
suppose that neuron $i$ receives an inhibitory (excitatory) connection
from neuron $j$. When the presynaptic neuron $j$ emits a spike at time
$t$, the potential of the postsynaptic neuron $i$ is decreased
(increased) by $J$ at time $t+\delta$ and returns exponentially to the
resting potential in a time $\tau$ which represents the integration
time constant of the membrane. In this simple model, the single time
$\delta$ is meant to represent the transmission delays but also and
most importantly, the longer time needed to obtain the full
hyperpolarization of the post-synaptic neuron corresponding to a given
presynaptic spike. Therefore, finding the correspondence between
$\delta$ and the different synaptic time scales of a more realistic
description needs some care. As pictorially shown in Fig.~\ref{IPSP},
$\delta$ should roughly be identified to the characteristic duration
of the synaptic currents. In the following, we thus refer to $\delta$,
which plays a crucial role in the generation of global oscillations,
as the "synaptic time".  The correspondence between $\delta$ and the
different synaptic time scales of a more realistic description is
further elaborated in Section \ref{section:synaptic} where synaptic
current of finite duration are considered.

Mathematically, the depolarization $V_i(t)$ of neuron $i$
($i=1,\ldots,N$) at its soma obeys the equation, 
\beq \tau
\dot{V}_i(t) = -V_i(t) + R I_i(t)
\label{potdyn}
\eeq 
where $I_i(t)$ are the synaptic currents arriving at the soma.
These synaptic currents are the sum of the contributions of spikes
arriving at different synapses (both local and external).  These spike
contributions are modelled as delta functions in our basic IF model:
\beq
\label{ispikes}
R I_i(t)= \tau \sum_j J_{ij} \sum_k \delta(t - t_j^k-\delta) 
\eeq
where the first sum on the r.h.s is a sum on different synapses
($j=1,\ldots,C+C_{ext}$), with postsynaptic potential (PSP) amplitude
(or efficacy) $J_{ij}$, while the second sum represents a sum on
different spikes arriving at synapse $j$, at time $t=t_j^k +\delta$,
where $t_j^k$ is the emission time of $k$-th spike at neuron $j$.  For
simplicity, we take PSP amplitudes equal at each synapse, i.e.~$J_{ij}
= J_{ext}>0$ for excitatory synapses and $J_{ij} =- J$ for inhibitory
ones.  External synapses are activated by independent Poisson
processes with rate $\nu_{ext}$.

A firing threshold $\theta$, completes the description of the IF
neuron : when $V_i(t)$ reaches $\theta$, an action potential is
emitted by neuron $i$, and the depolarization is reset to $V_r<\theta$
after a refractory period $\tau_{rp}$ during which the potential is
insensitive to stimulation. A typical value would be $\tau_{rp}\sim
2$ms.  We are interested here in network states in which the frequency
is much lower than the corresponding maximal frequency
$1/\tau_{rp}\sim 500$Hz.  In this regime, we have checked that the
exact value of $\tau_{rp}$ does not play any role.  Thus in the
following we set $\tau_{rp}$ to zero, for the sake of simplicity.

\begin{figure}
\setlength{\unitlength}{1cm}
\begin{picture}(15,20)
\put(1.,0){
\put(-3,-1.5){\includegraphics{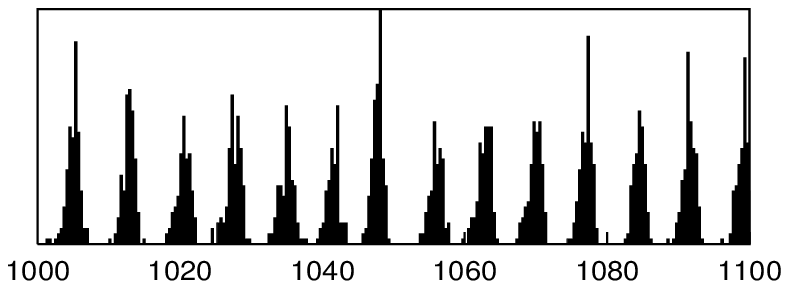}}\put(1,3){LFP}
\put(-3,5.75){\includegraphics{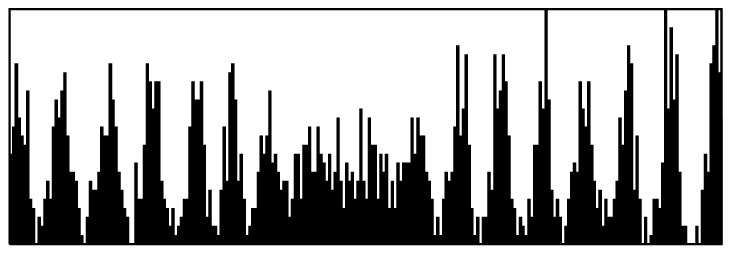}}\put(1,10.25){LFP}
\put(-3,13.){\includegraphics{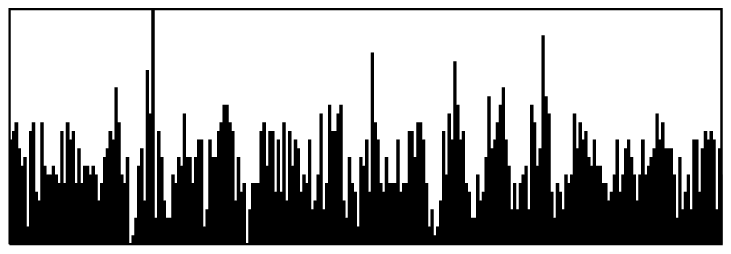}}\put(1,17.5){LFP}
\put(-3,1.3){\includegraphics{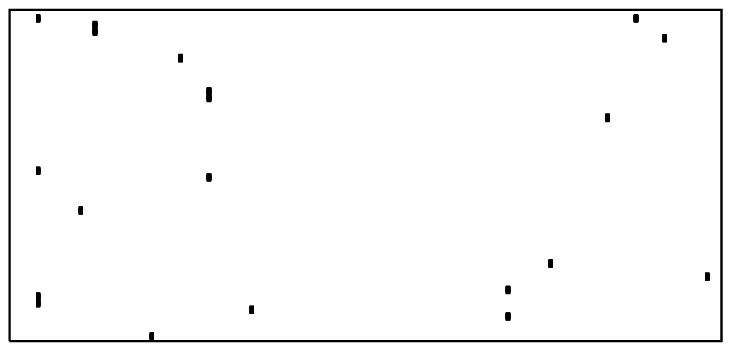}}
\put(-3,8.55){\includegraphics{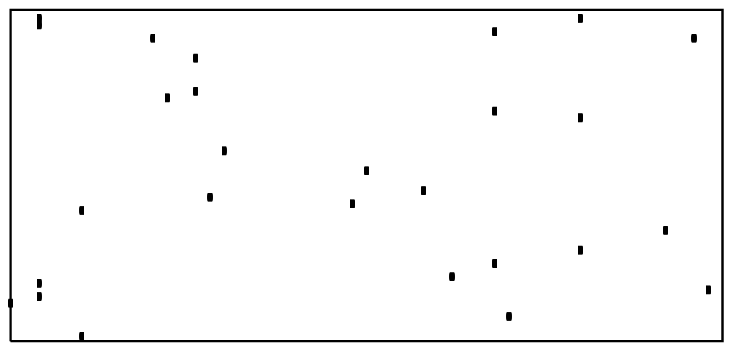}}
\put(-3,15.8){\includegraphics{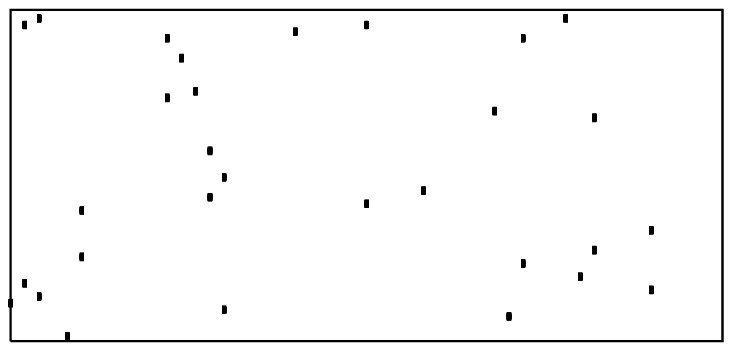}}
\put(7,-1.5){\includegraphics{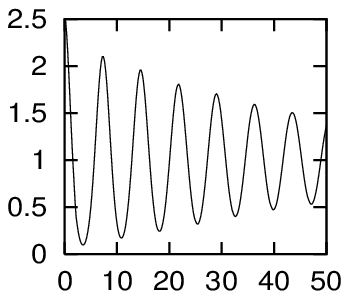}}\put(12.5,3){AC}
\put(7,5.75){\includegraphics{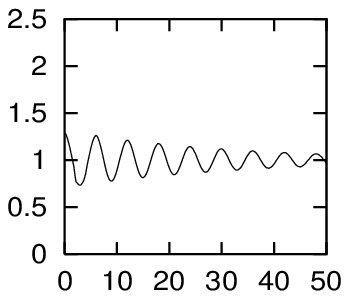}}\put(12.5,10.25){AC}
\put(7,13.){\includegraphics{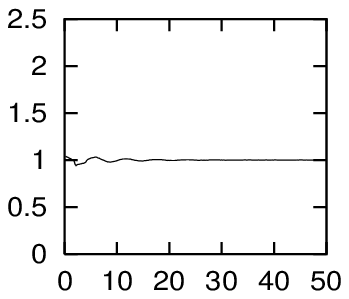}}\put(12.5,17.5){AC}
\put(7,2.2){\includegraphics{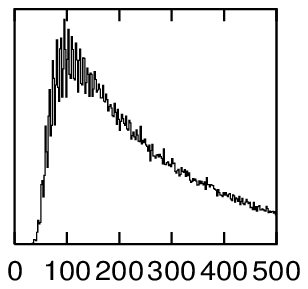}}\put(12.5,6.7){ISI}
\put(7,9.45){\includegraphics{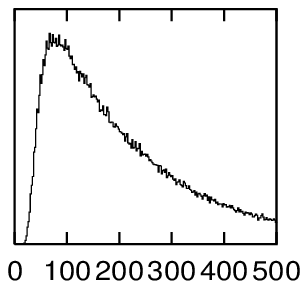}}\put(12.5,13.95){ISI}
\put(7,16.7){\includegraphics{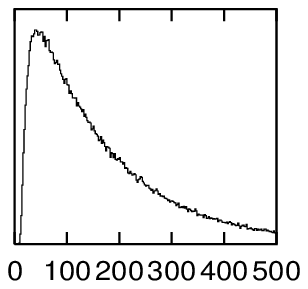}}\put(12.5,21.2){ISI}
\put(4.,-0.1){time(ms)}
\put(11.2,-0.1){time(ms)}
}
\put(0,21){A}
\put(0,13.75){B}
\put(0,6.5){C}
\end{picture}
\caption{Left: Time evolution of the global activity (LFP) during a
100ms interval of the dynamics of a network of 5,000 neurons (total
number of firing neurons in 0.4ms bins), together with spike rasters
of 50 neurons, for different values of the external noise:
$\sigma_{ext}=5$mV (A), 2.5mV (B), and 1 mV (C). Right:
autocorrelation of the global activity (AC) and inter-spike interval
(ISI) histogram averaged over 1000 neurons, corresponding to the left
pictures.  Note the different time scales of AC and ISI in abscissa.
Parameters: $\theta=20$mV, $V_r=10$mV, $\tau=20$ms, $\delta=2$ms,
$C=1000$, $J=0.1$mV, $\mu _{ext}=25$mV.}
\label{figuresim1}
\end{figure}

The outcome of a typical simulation is shown in
Figs.~\ref{figuresim1}.  Neurons are driven by the random external
excitatory input above threshold; however, since feedback interactions
are inhibitory, the global activity stays at rather low levels (about
5Hz for the parameters indicated in Fig.~\ref{figuresim1}). For weak
external noise levels ($\sigma_{ext}=1$mV), the global activity (total
number of firing neurons in 0.4ms bins) is strongly oscillatory with a
period of about 7 ms, as testified by Fig.~\ref{figuresim1}C.  On the
other hand, increasing the external noise level strongly damps and
decreases the amplitude of the global oscillation. Note that the
global activity should roughly correspond to the local field potential
(LFP) often recorded in neurophysiological experiments.  On the other
hand, even when the global activity is strongly oscillatory,
individual firing is extremely irregular as shown in the rasterfile of
50 neurons, Fig.~\ref{figuresim1}C (above the LFP), and in the
inter-spike interval histogram (to the right of the spike rasters).
In each oscillatory event only a small fraction of the neurons fire.

This oscillatory collective behavior is also shown by fast
oscillations in the temporal autocorrelation (AC) of the global
activity which are damped on a longer time scale
(Fig.~\ref{figuresim1}, to the right of the LFP).  It is also
reflected in the cross-correlations (CC) between the spike trains of a
pair of neurons, which are typically equal to the AC of the global
activity.

These simulation results raise several questions on the origin and
characteristics of the observed oscillations. What is the mechanism of
the fast oscillation?  In which parameter region is the network
oscillating?  What are the network parameters which control the
amplitude and the different time scales (frequency, damping time
constant) of the global oscillation? How do they scale with the
network size? The model is simple enough and an analytical study gives
precise answers to these questions as shown in the following sections.
 
\section{An analysis of the network dynamics}

\label{sec:an:analysis}

Several features simplify the analysis as noted in a previous study
(Amit and Brunel 1997a) of the neuron mean firing rates. First, as a
consequence of the network sparse random connectivity ($C\ll N$), two
neurons share a small number of common inputs and pair correlations
can be neglected in the limit $C/N \rightarrow 0$.  Second, we
consider a regime where individual neurons have a firing rate $\nu$
low compared to their inverse integration time $1/\tau$ and receive a
large number of inputs per integration time $\tau$, each input making
a contribution small compared to the firing threshold ($J \ll 
\theta$)\footnote{Typical numbers in cortex are $C=5000$, $\tau=20$ms,
$\nu=5$Hz, $J=0.1$mV, $\theta=20$mV so that $C\nu\tau$ is typically
several hundreds while $\theta/J$ is of order 100 (Abeles 1991,
Braitenberg and Shutz 1991). In the simulation shown in
Fig.~\ref{figuresim1} $C\nu\tau\sim 100$, $\theta/J\sim 200$.}.  In
this situation, the synaptic current of a neuron can be approximated
by an average part plus a fluctuating gaussian part, and the spike
trains of all neurons in the network can be self consistently
described by Poisson processes with a common instantaneous firing rate
$\nu(t)$ but otherwise uncorrelated from neuron to neuron (that is,
between $t$ and $t+dt$, a spike emission has a probability $\nu(t) dt$
of occurring for each neuron but these events occur statistically
independently in different neurons)
 
The  synaptic current at the soma of a neuron (neuron $i$) 
can thus be written as, 
\beq
R I_i(t) = \mu (t) + \sigma\sqrt{\tau} \eta_i(t)
\label{idiffusion}
\eeq
The average part $\mu (t)$ is  related to the firing rate 
at time $t-\delta$ and is a sum of local and external inputs
\beq
\mu = \mu _l + \mu _{ext} \,\,\mbox{ with }\,\,
\mu _l = - CJ\nu(t-\delta)\tau, \;\;\;\; \mu _{ext}= C_{ext}J_{ext}\nu_{ext}\tau
\label{mu}
\eeq
Similarly the fluctuating part,
$\sigma\sqrt{\tau} \eta_i(t)$, is given by the fluctuation in the sum of
internal and external poissonian inputs of rate $C\nu$
and $C_{ext}\nu_{ext}$. Its magnitude is given by
\beq
\sigma^2 = \sigma^2_l+\sigma^2_{ext} \,\,\mbox{ with }\,\,
\sigma_l = J\sqrt{C\nu(t-\delta)\tau}, \;\;\;\; \sigma_{ext}= 
J_{ext}\sqrt{C_{ext}\nu_{ext}\tau}
\label{sigma}
\eeq
and $\eta_i(t)$ is a gaussian white noise uncorrelated from neuron
to neuron,
$\langle \eta_i(t)\rangle =0$
and $\langle \eta_i(t) \eta_j(t')\rangle = \delta_{i,j}\delta(t-t')$.

Before describing our precise results, it may be useful to give simple
estimates which show how the neuron firing rates, the collective
oscillation frequency and the oscillatory threshold can be obtained
from Eqs.(\ref{idiffusion}-\ref{sigma}).

Let us first consider the stationary case. The case of
interest corresponds to $\mu <\theta$. When expression
(\ref{idiffusion}) is used for the synaptic current, the dynamics of
the neuron depolarization (\ref{potdyn}) is  a stochastic motion
in the harmonic potential $(V-\mu )^2$ truncated at the firing
threshold $V=\theta$. The neuron firing rate $\nu_0$ is  the
escape rate from this potential. For a weak noise, it is given by the
inverse of the time scale of the motion $1/\tau$ diminished by an
Arrhenius activation factor. So, one obtains the simple estimate (up
to an algebraic prefactor),
\beq
\nu_0\sim\frac{1}{\tau}\exp\left(-\frac{(\theta-\mu )^2}{\sigma^2}\right)
\label{simpstatrate}
\eeq
This becomes a self-consistent equation for $\nu_0$ once $\mu $ and
$\sigma$ are expressed in terms of $\nu_0$ using
Eq.~(\ref{mu},\ref{sigma}). The simple estimate (\ref{simpstatrate})
is made precise below by following Kramers's classic treatment of the
thermal escape over a potential barrier (Chandrasekhar 1943).

The origin of the collective oscillation can also be simply
understood.  An increase of activity in the network due to a
fluctuation provokes an increase in the average feedback inhibitory
input. Thus after a period of about one synaptic time the activity
should decrease due to the increase of the inhibitory input. This
decrease will itself provoke a decrease in the inhibitory input, and a
corresponding increase in the activity after a new period equal to the
synaptic time.  This simple argument 
 predicts a global oscillation
period of about a couple of times the synaptic time $\delta$, not too
far from the period observed in the simulations. However, it does not
seem to have been noted previously that a global oscillation
of period $\delta$ can in fact occur only if it is not masked by the
intrinsic noise in the system. The resulting oscillation threshold can
be simply estimated in the limit where $\delta$ is short compared to
the time scale of the depolarization dynamics.  During a short time
interval $\delta$, a neuron membrane potential receives from the local
network an average input of magnitude $C\nu_0 \delta J$. The
fluctuation in its membrane potential in the same time interval (due
to intrinsic fluctuations in the total incoming current) is $\sigma
\sqrt{\delta/\tau}$. The change in the average local input can be
detected only if it is larger than the intrinsic potential
fluctuations. A global oscillation can therefore occur only when
$$\frac{CJ\nu_0\tau}{\sigma} = -\frac{\mu _l}{\sigma}\,\stackrel{>}{\sim}\,
\sqrt{\frac{\tau}{\delta}}.$$

These simple estimations are confirmed by the analysis presented below
and replaced by precise formulas.

\subsection{Dynamics of the distribution of neuron potentials}

When pair correlations are neglected, the system can be described by
the distribution of the neuron depolarization $P(V,t)$, i.e.~the
probability of finding the depolarization of a randomly chosen neuron
at $V$ at time $t$. This distribution is  the (normalized)
histogram of the depolarization of all neurons at time $t$ in the
large $N$ limit $N\rightarrow\infty$. The stochastic equation
(\ref{potdyn},\ref{idiffusion}) for the dynamics of a neuron
depolarization can be transformed into a Fokker-Planck equation
describing the evolution of their probability distribution
(Chandrasekhar 1943)
\beq
\label{fp}
\tau\frac{\partial P(V,t)}{\partial t} = \frac{\sigma^2(t)}{2}
\frac{\partial^2 P(V,t)}{\partial V^2}+ \frac{\partial}{\partial
V}\left[ (V-\mu (t)) P(V,t)\right] \eeq 
The two terms in the r.h.s.~of
(\ref{fp}) correspond respectively to a diffusion term coming from the
current fluctuations and a drift term coming from the average part of
the synaptic input.  $\sigma(t)$ and $\mu (t)$ are related to
$\nu(t-\delta)$, the probability per unit time of spike emission at
time $t-\delta$, by Eq.~(\ref{mu},\ref{sigma}).  Note that the
Fokker-Planck equation has been used previously in studies of globally
coupled oscillators (Sakaguchi et al 1988, Strogatz and Mirollo 1991,
Abbott and van Vreeswijk 1993, Treves 1993).

The resetting of the potential at the firing threshold ($V=\theta$)
imposes the absorbing boundary condition $P(\theta,t)=0$. Moreover,
the probability current through $\theta$ gives the probability of
spike emission at $t$,
\beq
\frac{\partial P}{\partial V}(\theta,t) =  -\frac{2\nu(t)\tau}{\sigma^2(t)}
\label{bcth}
\eeq

At the reset potential
$V=V_r$, $P(V,t)$ is continuous but the entering probability current
imposes the following derivative discontinuity,
\beq
\frac{\partial P}{\partial V}(V_r^+,t) - 
\frac{\partial P}{\partial V}(V_r^-,t) =  -\frac{2\nu(t)\tau}{\sigma^2(t)}
\label{bcv}
\eeq

At $V= -\infty$, 
$P$ should tend sufficiently quickly toward zero to be integrable, i.e.
\beq
\lim_{V\rightarrow -\infty} P(V,t)=0 \;\;\; 
\lim_{V\rightarrow -\infty} V P(V,t)=0.
\label{bci}
\eeq

Last, $P(V,t)$ is a probability distribution and should satisfy
the normalization condition
\beq
\int_{-\infty}^{\theta} P(V,t) dV = 1
\label{norm}
\eeq

\subsection{Stationary states}

\begin{figure}
\setlength{\unitlength}{1cm}
\begin{picture}(14,6)
\put(0,-1){\includegraphics{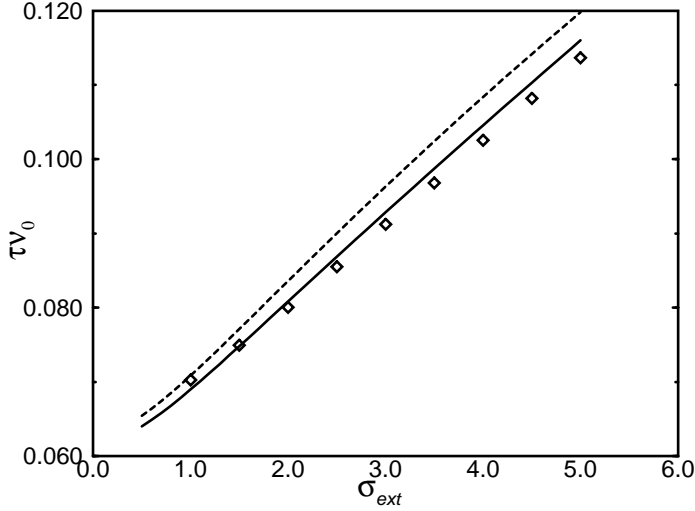}}
\end{picture}
\caption[]{The neuron firing rate vs $\sigma_{ext}$: simulation
 ($\diamond$); solution of Eq.~(\ref{kraex})(full line); solution of
 the approximate asymptotic form (\ref{kraas}) (dashed line). Others
 parameters are fixed as in Fig.~2 : $\tau=20$ms, $J=0.1$mV, $C=1000$,
 $N=5000$, $\theta=20$mV, $V_r=10$mV, $\mu _{ext}=25$mV, $\delta=2$ms.}
\label{nurates}
\end{figure}

We first consider stationary solutions $P(V,t)=P_0(V)$. Time independent
solutions of
Eq.~(\ref{fp}) satisfying the
boundary conditions (\ref{bcth},\ref{bcv},\ref{bci}) are given by
\beq
\label{statdistr}
P_0(V) =
2\frac{\nu_0\tau}{\sigma_0} \exp\left(-\frac{(V-\mu _0)^2}{\sigma_0^2}
\right)\int_{\frac{V-\mu _0}{\sigma_0}}^{\frac{\theta-\mu _0}{\sigma_0}}
\Theta\left(u - \frac{V_r-\mu _0}{\sigma_0}\right) e^{u^2}du
\label{p0}
\eeq
with
\beq
\mu _0 =-CJ\nu_0 \tau +\mu _{ext}, \;\;\;\;\sigma_0^2 = CJ^2\nu_0\tau +
\sigma_{ext}^2
\eeq
(in (\ref{p0}), $\Theta(x)$ denotes the Heaviside function, 
$\Theta(x)=1\,$ for $x>0$ and $\Theta(x)=0$ otherwise).
The normalization condition (\ref{norm}) provides the self-consistent
condition which determines
$\nu_0$ 
\beqa
\frac{1}{\nu_0\tau} &=& 2 
\int_{\frac{V_r-\mu _0}{\sigma_0}}^{\frac{\theta-\mu _0}{\sigma_0}}
du e^{u^2}\int_{-\infty}^{u}dv e^{-v^2}
\nonumber\\
&=& \int_0^{+\infty} du e^{-u^2}\left[\frac{e^{2y_{\theta} u}
-e^{2y_r u}}{u}
\right]
\label{kraex}
\eeqa
with $y_{\theta}=\frac{\theta-\mu _0}{\sigma_0}, y_r=
\frac{V_r-\mu _0}{\sigma_0}$.
In the regime $(\theta-\mu _0)\gg \sigma_0$, Eq.~(\ref{kraex}) becomes
\beq
\nu_0\tau\simeq\frac{(\theta-\mu _0)}{\sigma_0\sqrt{\pi}}
\exp\left(-\frac{(\theta-\mu _0)^2}{\sigma_0^2}\right)
\label{kraas}
\eeq
In Fig.~(\ref{nurates}), the firing rates obtained by solving
Eq.~(\ref{kraex}) and (\ref{kraas})
are compared with those obtained from simulations of the 
network. It shows an almost linear increase in the rates as a function
of $\sigma_{ext}$ in the
range 3-6Hz and a good agreement between Eq.~(\ref{kraex})
and the results of simulations. The asymptotic expression (\ref{kraas})
is also rather close to the simulation results in this range of $\sigma$.

\subsection{Linear stability of the stationary states}

We can now investigate in which parameter regime the time independent
solution $(P_0(V),\nu_0)$ is stable. 
To simplify the study  of the Fokker-Planck equation (\ref{fp}), it is
convenient to
rescale $P$, $V$
and $\nu$ by 
\beq
P=\frac{2\tau\nu_0}{\sigma_0}Q,\,\,\, 
y=\frac{V-\mu _0}{\sigma_0},\,\,\,
\nu=\nu_0(1+n(t))
\label{pytrntext}
\eeq
$y$ is the difference between the membrane potential and
the average input in the stationary state, in units of the average fluctuation of the input in the stationary state. 
$n(t)$ corresponds to the relative variation of the instantaneous frequency
around the stationary frequency.
After these rescalings, Eq.(\ref{fp})
becomes
\beq
\tau \frac{\partial Q}{\partial t}=\frac{1}{2} \frac{\partial^2 Q}{\partial y^2} + 
\frac{\partial}{\partial y}(yQ) + n(t-\delta)
\left(G \frac{\partial Q}{\partial y} +\frac{H}{2} 
\frac{\partial^2 Q}{\partial y^2}\right),
\label{Eq}
\eeq
where $G$ is the ratio between the mean local
inhibitory inputs and $\sigma_0$, and $H$ is the ratio between the variance
of the local inputs and the total variance (local plus external):
\beq
G=\frac{C J\tau\nu_0}{\sigma_0}=\frac{-\mu _{0,l}}{\sigma_0},\,\,\,
H=\frac{C J^2\tau\nu_0}{\sigma_0^2}=\frac{\sigma_{0,l}^2}{\sigma_0^2},
\label{ghtext}
\eeq
These parameters are a measure of the relative strength of the recurrent
inhibitory interactions.

Eq.~(\ref{Eq}) holds on the two intervals $-\infty<y<y_r$ and 
$y_r<y<y_{\theta}$.
The boundary conditions on $Q$ are imposed at
$y_{\theta}=\frac{\theta-\mu _0}{\sigma_0}$ and
$y_r=\frac{V_r-\mu _0}{\sigma_0}$. Those on the derivatives of $Q$ read,
\beq
\frac{\partial Q}{\partial y}(y_{\theta},t)=
\frac{\partial Q}{\partial y}(y_r^+,t)-\frac{\partial Q}{\partial y}(y_r^-,t)
=-\,\frac{1+n(t)}{1+H n(t-\delta)}
\eeq

The linear stability of the stationary solution is studied in detail
in Appendix \ref{app:linear:stability}. This can be done in a
standard way (Hirsch and Smale, 1974) by expanding $Q=Q_0+Q_1+\ldots$ and
$n=n_1+\ldots$ around the steady state solution.  The linear equation
obtained at first order has solutions which are exponential in time,
$Q_1=\exp(w t/\tau)\hat Q_1$, $n_1\sim \exp(w/\tau) \hat{n}_1$, where
$w$ is a solution of the eigenvalue equation (\ref{eigeneq}) of the
Appendix.  The stationary solution becomes unstable when the real part
of $w$ becomes positive.

When the synaptic time $\delta$ becomes much smaller than $\tau$, the roots
$w$ of this equation become large. We consider the regime
$\delta/\tau\ll 1$ but $\delta/\tau\gg 1/C$, which is the relevant
case in simulations and correspond to the realistic regime.
$\delta/\tau\gg 1/C$ is needed because otherwise the equations giving
$G$ and $H$ become inconsistent with the condition $\tau\nu_0\ll 1$.
At the oscillatory instability onset, $w$ is purely imaginary
$w=i\omega_c$, where $\omega_c/\tau$ is the frequency of the
oscillation which develops.  The eigenvalue equation takes in the
limit $\delta/\tau\rightarrow 0$, $\omega\rightarrow \infty$ the form
\beq
[\frac{G}{\sqrt{\omega_c}}(i-1)+H]\, \exp(-i \omega_c \delta/\tau) =1.
\eeq
In this limit, the instability line in the parameter space $(G,H)$ is obtained
parametrically as
\begin{eqnarray*}
G&=&\sqrt{\omega_c}\sin\left(\frac{\omega_c\delta}{\tau}\right) \\
H&=&\sin\left(\frac{\omega_c\delta}{\tau}\right) + \cos
\left(\frac{\omega_c\delta}{\tau}\right)
\end{eqnarray*}
$H$ is by definition constrained to be between 0 and 1
(it is the ratio between local and total variances): $H=0$ corresponds
to the limit of very large external fluctuations, $\sigma_{ext}\gg \sigma_l$,
while $H=1$ corresponds to $\sigma_{ext}=0$.
We find that the frequency of the oscillation varies from 
\begin{eqnarray}
\frac{\omega_c}{\tau} & = & \frac{3\pi}{4\delta}\quad\mbox{ when } H=0,  \mbox{ to } 
\nonumber \\
\frac{\omega_c}{\tau} & = & \frac{\pi}{2\delta}\quad\mbox{ when } H=1. 
\label{omega:limit}
\end{eqnarray}
This corresponds to an oscillation with a period between $8\delta/3$ and
$4\delta$, not too far from the value $2\delta$ obtained by 
simple arguments. At the same
time the critical value of $G$ goes from 
\begin{eqnarray*}
G_c & = & \sqrt{\frac{3\pi\tau}{8\delta}}\quad\mbox{ when } H=0, \mbox{ to } \\
G_c & = & \sqrt{\frac{\pi\tau}{2\delta}}\quad\mbox{ when } H=1. 
\end{eqnarray*}
Again we find that it is proportional to $\sqrt{\tau/\delta}$ as anticipated.

\begin{figure}
\setlength{\unitlength}{1cm}
\begin{picture}(14,4.5)
\put(6.5,0){\put(-2,-1.5){\includegraphics{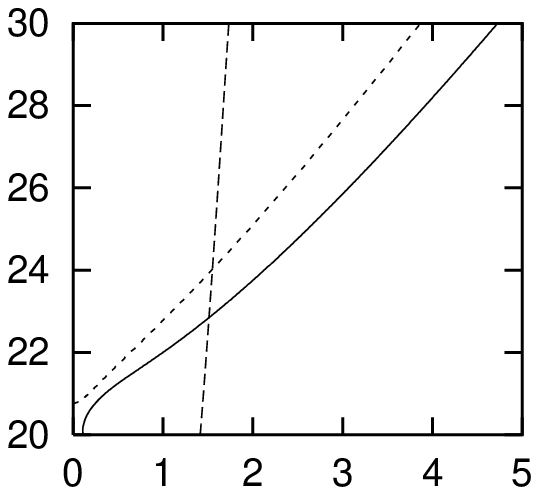}}
\put(0.5,3.3){$\mu _{ext}$}
\put(4.5,-0.2){$\sigma_{ext}$}}
\put(-0.5,0){\put(-2,-1.5){\includegraphics{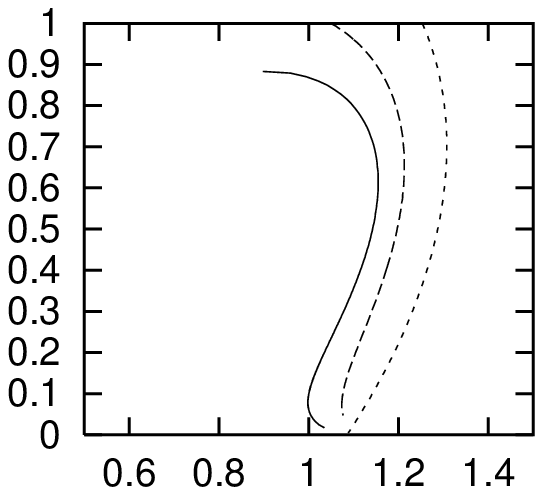}}
\put(0.2,3.3){$H$}
\put(4.,-0.2){$G\sqrt{\delta/\tau}$}}
\put(2.5,3.5){SS}\put(5.7,3){OS}\put(9,4.5){OS}\put(12,2.5){SS}
\end{picture}
\caption{Left: instability line in the plane
$(H,G\sqrt{\delta/\tau})$.  Full line: instability line for parameters
of Fig.\ref{figuresim1}, and $\delta=0.1\tau$.  Long-dashed line:
$\delta=0.05\tau$. Short-dashed line: asymptotic limit
$\delta/\tau\rightarrow 0$. The stationary state (SS) is unstable to
the right of the instability line, where an oscillatory instability
develops (OS).  Right: instability line in the plane
$(\mu _{ext},\sigma_{ext})$.  Full line: parameters of Fig.2, and
$\delta=0.1\tau$.  Short-dashed line is constructed taking the
asymptotic instability line in the plane $(H,G\sqrt{\delta/\tau})$,
and calculating the corresponding instability line in
$(\mu _{ext},\sigma_{ext})$ with $\delta=0.1\tau$. The stationary state
(SS) becomes unstable above the instability line. The long dashed line
shows the average ($\mu _{ext}$) and the fluctuations ($\sigma_{ext}$)
of the external inputs when the frequency of a Poissonian external
input through synapses of strength $J_{ext}=0.1$mV is varied.  For low
external frequencies the network is in its stationary state. When the
external frequency increases the network goes to its oscillatory state
(OS).}
\label{musi}
\end{figure}

This instability line can be translated in terms of the parameters
$\mu _{ext}$, $\sigma_{ext}$, and calculated numerically using
Eq.~(\ref{eigeneq}) for any value of the network parameters.  This
line of instability in the plane $(\mu _{ext},\sigma_{ext})$ is shown
in the right part of Fig.~\ref{musi}. The stationary solution is
unstable above the full line. Thus, if the external input is
Poissonian, an increase in the frequency of external stimulation will
typically bring the network from the stationary to the oscillatory
regime, as indicated by the dashed line in Fig.~\ref{musi}, which
represents the average ($\mu _{ext}$) and the fluctuations
($\sigma_{ext}$) of the external inputs when the frequency of a
Poissonian external input through synapses of strength $J_{ext}=0.1$mV
is varied.

\subsection{Weakly non-linear analysis}

\label{sec:weakly:nl}
The linear stability analysis of the previous section shows that a small
oscillation grows
when one crosses the instability line in the plane $\mu _{ext}$,
$\sigma_{ext}$. But it does not say much on the resulting
characteristics of the resulting finite amplitude oscillation.
In order to describe it and to be able
to quantitatively compare analytic results to
simulation data, one needs to compute the  non linear
terms which saturate the instability growth. This can be
done in a standard manner (Bender and Orszag, 1987)
by computing terms beyond the linear
order in an expansion around the stationary state.
The explicit computation is detailed in
Appendix \ref{app:weakly:nl}. The collective oscillation is determined
by the deviation $n_1$ of the neuron firing rate from its stationary
value:
$$
n_1(t) = \hat n_1(t) \exp(i\omega_c t/\tau) + \hat n_1^{\star}(t) \exp(-i\omega_c t/\tau)
$$
$\hat n_1$ determines the amplitude of the collective oscillation as
well as the nonlinear contribution to its frequency in the vicinity of
the instability line.

The analysis shows that the dynamics of the (small) deviation
around the stationary firing rate can be described by the reduced
equation
\beq
\label{eqmotiontext}
\tau \frac{d\hat n_1}{dt} = A \hat n_1 - B |\hat n_1|^2 \hat n _1 \eeq
in which $A$ and $B$ are complex numbers. The value of $A$ comes
 from the linear stability analysis. If Re$(A) <0$ a small initial value
of $n_1$ decays and the stationary state is stable. On the contrary, if
$Re(A)>0$ a global oscillation develops. When $|\hat n_1|$ grows, the second
nonlinear term on the r.h.s. of (\ref{eqmotiontext}) becomes important.
It is found
here that Re$(B)>0$ (a "normal" or "supercritical" Hopf bifurcation) so
that the nonlinear term saturates the linear growth. 
The
characteristics of the oscillatory final state comes from the balance
between the two terms.

The explicit expression of $A$ and $B$ is given in Eqs.~(\ref{a},\ref{b})
as a ratio of
hypergeometric functions of the network parameters.
$A$ depends linearly on the deviation of the
parameters $G$ and $H$ from their critical values, i.e.~$G-G_c$,
$H-H_c$.  In the limit $\delta/\tau\rightarrow 0$, the expressions of
$A$ and $B$ simplify.  For example, when $H=0$ (large external
fluctuations), we find in the limit $\delta/\tau\rightarrow 0$ 
\beqa A
& = & \frac{\tau}{\delta}\frac{\left(1+2i/3\pi\right)}{(1+4/9\pi^2)}
\frac{G-G_c}{G_c} \simeq \frac{\tau}{\delta} \left(1.35 +
0.29i\right)\frac{G-G_c}{G_c} \nonumber \\ B & = &
\frac{\tau}{\delta}\left(\frac{9\pi^2}{4+9\pi^2}\right)\left[\frac{13-5\sqrt{2}}{10}-\frac{9-5\sqrt{2}}{15\pi}
+i\left(\frac{13-5\sqrt{2}}{15\pi}+\frac{9-5\sqrt{2}}{10}\right)\right]
\nonumber \\ & \simeq & \frac{\tau}{\delta} (0.53+0.30 i) 
\label{ABsimp}
\eeqa

Generally,
the complex numbers $A$ and $B$ can be written in terms of their
real and imaginary parts, $A=A_r +iA_i$, $B=B_r +iB_i$.
On the critical line, i.e. for $G=G_c$, $H=H_c$, $A_r=A_i=0$; 
above the critical line an 
instability develops, $A_r>0$, proportionnally to $G-G_c$ and
$H-H_c$. The amplitude of this instability
is controlled by the cubic term. The stable limit cycle
solution of Eq.~(\ref{eqmotiontext}),
above the critical line, is
\beq
\label{noiselessnu1text}
\hat n_1(t) = R \exp\left(i\Delta\omega \frac{t}{\tau}\right)
\eeq
where 
$$
R = \sqrt{\frac{A_r}{B_r}}\quad\mbox{ and } \quad\Delta\omega = A_i - B_i\frac{A_r}{B_r}
$$

The autocorrelation (AC) of the global activity, normalized by
$\nu_0$, is, when $A_r>0$, \beqa C(s) & = & \lim_{T\rightarrow\infty}
\frac{1}{T-s} \int_0^{T-s} (1+n_1(t))(1+n_1(t+s)) dt \\ \nonumber & =
& 1+ 2R^2 \cos\left[(\omega_c+\Delta\omega)s/\tau\right]\nonumber
\eeqa The AC is  a cosine function of frequency
$(\omega_c+\Delta\omega)/\tau$ and amplitude $R^2$. Compared with the
AC function observed in the simulation, Fig.~\ref{figuresim1}C, we see
a qualitative difference: there is no damping of the oscillation. The
next Section shows that the damping is due to finite size effects. We
analyze them before comparing quantitatively the analytical results
with simulations.

\subsection{Finite size effects and phase diffusion of the collective 
oscillation}

We discuss the effect of having a large but only finite number of
neurons in the network. It is well-known that for stochastic dynamics,
a sharp transition can only occur in the limit $N\rightarrow\infty$
and that it will be smoothened by finite size effects.  In the sparse
connectivity limit, which allows to treat the quenched random geometry
of the lattice in an annealed fashion\footnote{Here we do not consider
the correlations due to the quenched connectivity for finite
$\epsilon$. These correlations would give small corrections to the
parameters calculated in the limit $\epsilon \rightarrow $0, but do
not give rise to qualitatively new effects for the global activity
such as the phase diffusion phenomenon discussed in this section.} the
fluctuations in the input of a given neuron $i$ can be seen as the
result of the randomness of two different processes: the first is the
spike emission process $S(t)$ of the whole network; and the second,
for each spike emitted by the network, is the presence or absence of a
synapse between the neuron that emitted the spike and the considered
neuron: if a spike is emitted at time $t$, $\rho_i(t)=1$ with
probability $C/N$, and 0 otherwise. The input to the network is then
$$
RI_i(t) = -J\tau \rho_i(t) S(t-\delta)
$$
Both processes can be decomposed between their mean and their fluctuation,
$$
\rho_i(t)=\frac{C}{N}+\delta \rho_i(t),\quad
S(t)= N\nu(t) + \delta S(t)
$$
Thus the input becomes
$$
RI_i(t) = \mu (t) - J\tau N\nu(t) \delta \rho_i(t)  -J\tau\frac{C}{N} \delta 
S(t)
$$
in which $\mu (t)$ is given by Eq.~(\ref{mu}).
The input is 
the sum of a constant part $\mu $, and of two distinct random processes
superimposed on $\mu $: the first is uncorrelated from neuron to neuron,
and we have already seen in Section \ref{sec:an:analysis} 
that it can be described
by $N$ uncorrelated 
Gaussian white noises $\sigma\sqrt{\tau}\eta_i(t)$, $i=1,\ldots,
N$ where
$<\eta_i(t)\eta_j(t')> =\delta_{ij}\delta(t-t')$.
The second part is independent of $i$: it comes from the intrinsic
fluctuations in the spike train of the whole network which are
seen by all neurons. This part becomes negligible when $\epsilon=C/N\rightarrow
0$, but can play a role as we will see when $C/N$ is finite.
The 
global activity in the network is essentially a Poisson
process with instantaneous frequency $N\nu(t)$. Such a Poisson process
has mean $N\nu(t)$, which is taken into account in
$\mu $, and variance $N\nu(t)\delta(t-t')$.
The fluctuating part of this process is well approximated by
a Gaussian white noise $\sqrt{N\nu_0}\xi(t)$, where $\xi(t)$ satisfies
$<\xi(t)>=0$, $<\xi(t)\xi(t')>=\delta(t-t')$. Note that for simplicity
we take the variance of this noise to be independent of time, which
is the case for $ n_1(t)\ll 1$.
These fluctuations are global and perceived by all neurons in the network.
Thus, the 
mean synaptic
input received by the neurons becomes
$$
CJ\tau\nu(t) + J\sqrt{\epsilon C\nu_0\tau}\sqrt{\tau}\xi(t)+\mu _{ext}
$$
Inserting this mean synaptic input in the drift term
of the Fokker-Planck equation, we can rewrite Eq.~(\ref{Eq}) as
\beq
\tau \frac{\partial Q}{\partial t}=\frac{\partial}{\partial y}
\{[y+ G n(t-\delta) + \eta \sqrt{\tau}\xi(t)] Q\} +\frac{1}{2} 
\frac{\partial^2 Q}{\partial y^2}
\eeq
where $\eta$
denotes the intensity of the noise stemming from these global fluctuations.
$\eta$ tends to zero as the network size increases 
\beq
\label{eta}
\eta=\sqrt{\epsilon}\frac{\sigma_0^l}{\sigma_0}
\eeq

Taking into account this global 
noise term in the derivation of the reduced equation,
we obtain, after some calculations described in Appendix \ref{app:noise},
\beq
\label{noisymotiontext}
\tau\frac{d\hat n_1}{dt} = A \hat n_1 - B |\hat n_1|^2 \hat n_1 + D \sqrt{\tau} \zeta(t)
\eeq
in which $A$, $B$ and $D$ are given by Eqs.~(\ref{a},\ref{b},\ref{c}),
and $\zeta$ is a complex white noise
such that
$<\zeta(t)\zeta^{\star}(t')>=\delta(t-t')$. $D$ is proportional to $\eta$,
i.e.~to both the square root of the connection probability and to
the ratio between local and total fluctuations.

Thus, the effect of the finite size of the network is to add a small
stochastic component to the evolution equation of $n_1$, 
Eq.~(\ref{noisymotiontext}). Its main effect is to produce a phase diffusion
of the collective oscillation {\footnote This global phase diffusion in
a network of finite size is well-known (see e.g. (Rappel and Karma, 1996) for a simple example)} which leads to the damping of the oscillation
in the autocorrelation function.

\subsubsection*{Amplitude of the autocorrelation}

From the reduced Eq.~(\ref{noisymotiontext}), one can compute exactly
the autocorrelation at zero time $C(0)$ as shown
in Appendix \ref{app:noise}.
This gives :
\begin{itemize}
\item
In the stationary regime far from the critical line, $A_r <0,|D|/|A_r|\ll 1
$:
\beq
C(0) -1 \sim \frac{|D|^2}{|A_r|} \sim O\left(\frac{C}{N}\right)
\label{c01}
\eeq
The amplitude of the fluctuations in the global activity are proportional
to $C/N$ and thus vanish when the connection probability goes to zero.

\item
On the critical line, $A_r=0$
\beq
C(0)-1 = \frac{2 |D|}{\sqrt{\pi B_r}}\sim O\left(\sqrt{\frac{C}{N}}\right)
\label{c02}
\eeq
The amplitude of the fluctuations are proportional to the square root
of the connection probability.

\item
In the oscillatory regime far from the critical line,
$A_r >0, |D|/A_r \ll 1$ :
\beq
C(0)-1 \sim \frac{2A_r}{B_r}\sim O\left(1\right)
\label{c03}
\eeq
In this regime the amplitude of the oscillation is to leading
order independent of the noise amplitude.
\end{itemize}

\subsubsection*{Oscillations below the critical line}

In the stationary regime far from the critical line,
the fluctuations of activity $n_1$ provoked by the noise term
can be considered small and thus we can neglect the cubic term.
It is then easy to calculate the autocorrelation (AC) of the
activity,
\beq
\label{ACbelow}
C(s) = 1+ \frac{|D|^2}{|A_r|} \exp\left(-\frac{|A_r|s}{\tau}\right)
\cos\left([\omega_c + A_i]\frac{s}{\tau}\right)
\eeq
It is a damped cosine function. The damped oscillation
has frequency $(\omega_c +A_i)/\tau$ and  damping time constant
proportional to $\tau/|A_r|$. The amplitude of the autocorrelation
function is proportional to $C/N$.

\subsubsection*{Oscillations above the critical line}

In the oscillatory regime far from the critical line,
we find in Appendix \ref{app:noise}
an AC function of the form
\beq
\label{ACabove}
C(s) = 1 + 2\frac{A_r}{B_r} \cos\left((\omega_c+\Delta \omega) s/\tau\right)\exp\left(-\frac{\gamma^2(s)}{2}
\right)
\eeq
It is again a damped cosine function.
The damping factor $\exp\left(-\gamma^2(s)/2
\right)$ is different from an exponential only at short times
$s \sim \delta$. At longer times, $s\gg \delta$, we obtain again an 
exponential
$$
\exp\left(-\frac{\gamma^2(s)}{2}\right) = \exp\left(-\frac{|D|^2}{4 R^2}\left(1+
\frac{B_i^2}{B_r^2}\right)\frac{s}{\tau}\left[1+\frac{|D|^2}{2A_r}+
 O\left(|D|^4\right)\right]
\right)
$$
The damping time constant is proportional to leading order in $|D|$
to $1/|D|^2\sim
N/C$, i.e.~to the inverse of the connection probability. When $N$ goes to
infinity at $C$ fixed the `coherence time' of the oscillation increases
linearly with $N$.

This `phase diffusion' effect is the main  finite
size effect above the critical line. Both the amplitude and frequency
of the oscillation are essentially unaffected by these finite size effects.

\subsection{Comparison between simulations and theory}

The autocorrelation (AC) of the global activity was computed for each set
of parameters from a simulation of 20 seconds. Few longer
simulations were performed  as a check. The autocorrelation obtained in the
longer simulations are essentially identical to the one obtained
in the 20s simulation.

Since the analysis predicts AC functions described
by damped cosine functions, a least square
fit of all AC functions was performed with
such functions. Thus the full AC is reduced to three parameters,
its amplitude at zero lag $C_0$, its frequency $\omega$, and
its damping time constant (or coherence time) $\tau_c$
$$
C(s) = 1 + C_0 \exp\left(-\frac{|s|}{\tau_c}\right) \cos(\omega s)
$$
We then compared the result of the fitting procedure with the
analytical expressions.

We have varied the magnitude of the
external noise $\sigma_{ext}$ from 0 to 5 mV. This brings the network
from the `oscillatory' to the `stationary' state. 

\begin{figure}
\setlength{\unitlength}{1cm} 
\begin{picture}(14,20)
\put(0,14){\put(0,6.5){A}\put(-1,-1.5){\includegraphics{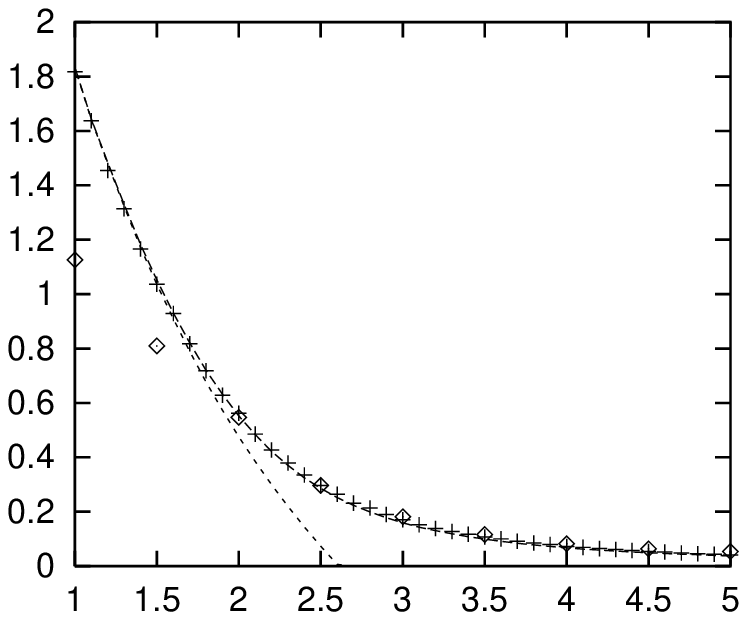}}
	\put(1,4){$C_0$}
}
\put(0,7){\put(0,6.5){B}
	\put(-1,-1.5){\includegraphics{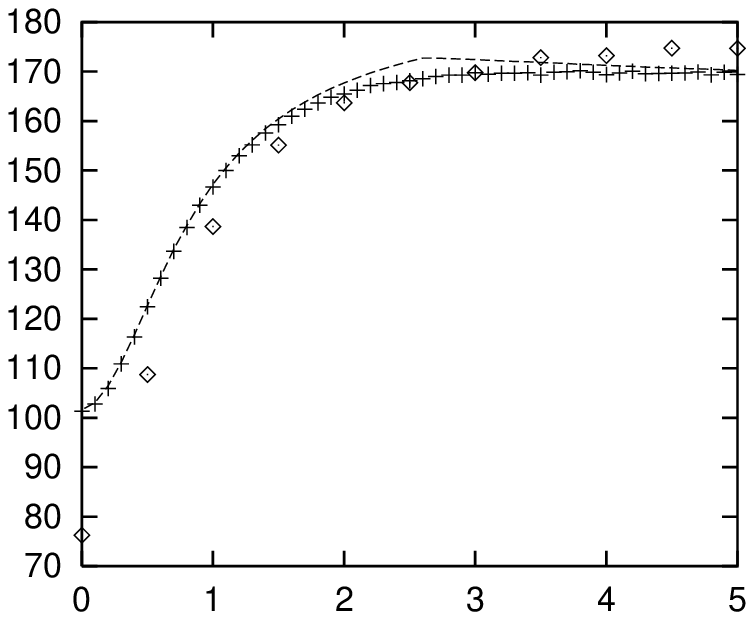}}
	\put(0.5,4){$\frac{\omega}{2\pi}$(Hz)}
}
\put(0,6.5){C}\put(-1,-1.5){\includegraphics{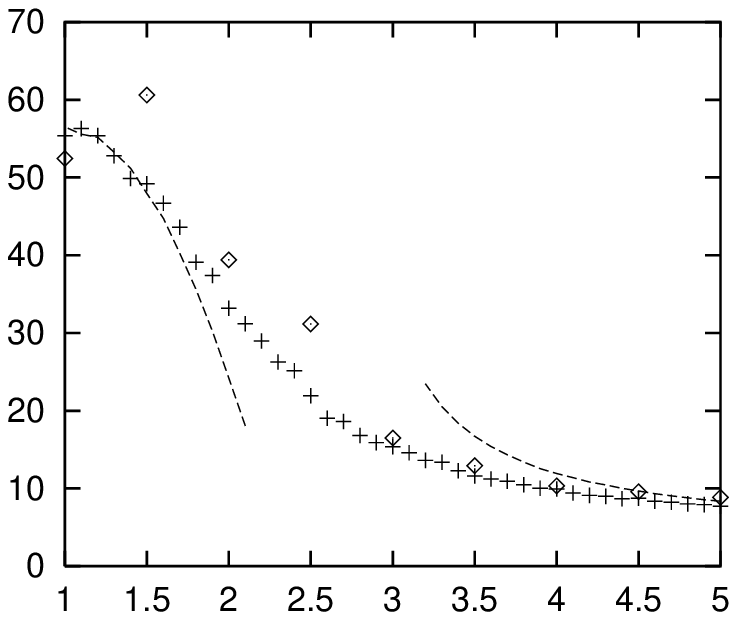}}
\put(6,0){$\sigma_{ext}$(mV)}
\put(0.5,4){$\tau_c$(ms)}
\end{picture}
\caption[]{Parameters of the AC function vs $\sigma_{ext}$.
A. Amplitude of the AC at zero lag.  B. Frequency. C. Damping time
constant.  Diamonds: simulation of the full network.  Crosses :
simulation of the reduced equation.  Dashed lines: theory.
In A, the short-dashed line represents the amplitude in the limit
$N\rightarrow\infty$
Parameters: $\tau=20$ms, $J=-0.1$mV, $C=1000$, $N=5000$,
$\theta=20$mV, $V_r=10$mV, $\mu _{ext}=25$mV, $\delta=2$ms.}
\label{siparams}
\end{figure}

In Fig.~\ref{siparams} we plot together
the results of simulations and theory. In these figures the diamonds 
are the simulation results; the dashed lines, the analytical results.
In A, the short-dashed line indicates the amplitude in the limit
$N\rightarrow\infty$, while the long-dashed line indicates the 
amplitude calculated analytically taking into account finite size effects.
Last, the crosses are obtained simulating numerically the reduced
equation, Eq.~\ref{noisymotiontext}. We find that, in the `stationary'
regime as well as in the oscillatory regime close to the bifurcation
point, the amplitude of the oscillation obtained in the simulation
is in very good agreement with the calculation (Fig.~\ref{siparams}.A). 
On the other hand,
as the amplitude of the oscillation becomes of the same order as the
average frequency, $C_0\sim 1$, higher order effects become important
and the calculation overestimates the amplitude of the AC.
For the frequency of the oscillation (Fig.~\ref{siparams}.B), the calculation
reproduces quite well the results of the simulations, except for
very low noise levels, for which we are rather far from the bifurcation
point. Note that the frequency ranges for this set of parameters from
70 to 180Hz, depending on the level of external noise. Thus, without
varying the time constants $\tau$ and $\delta$, we find that the same network
is able to sustain a collective oscillation at quite different frequencies.

Last, the approximate analytical expressions
for the damping time constant agree
well with the simulation away from the bifurcation point, as expected
(Fig.~\ref{siparams}.C).
On the other hand, the simulation of the reduced equation is in good
agreement with the network simulations in the whole range
of $\sigma_{ext}$.

\begin{figure}
\setlength{\unitlength}{1cm} 
\begin{picture}(14,20)
\put(0,14){\put(0,6.5){A}\put(-1,-1.5){\includegraphics{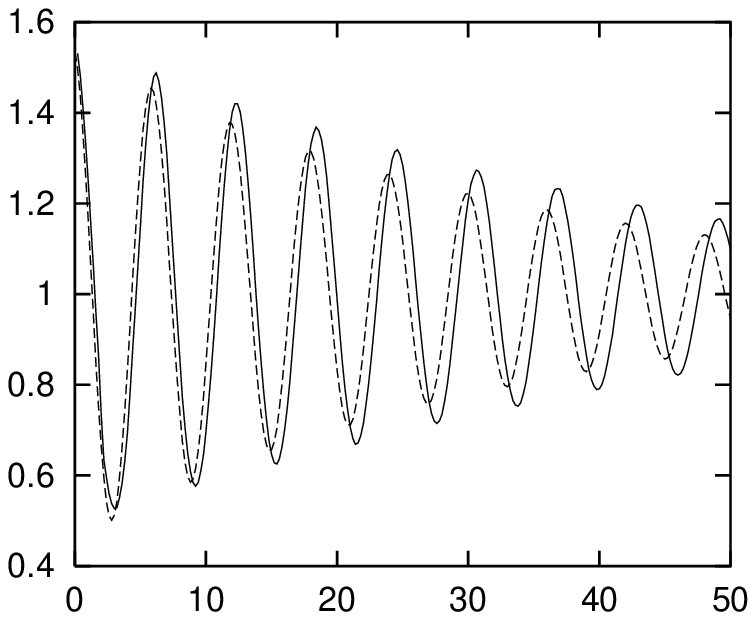}}
\put(1,4){$C(t)$}
}
\put(0,7){\put(0,6.5){B}\put(-1,-1.5){\includegraphics{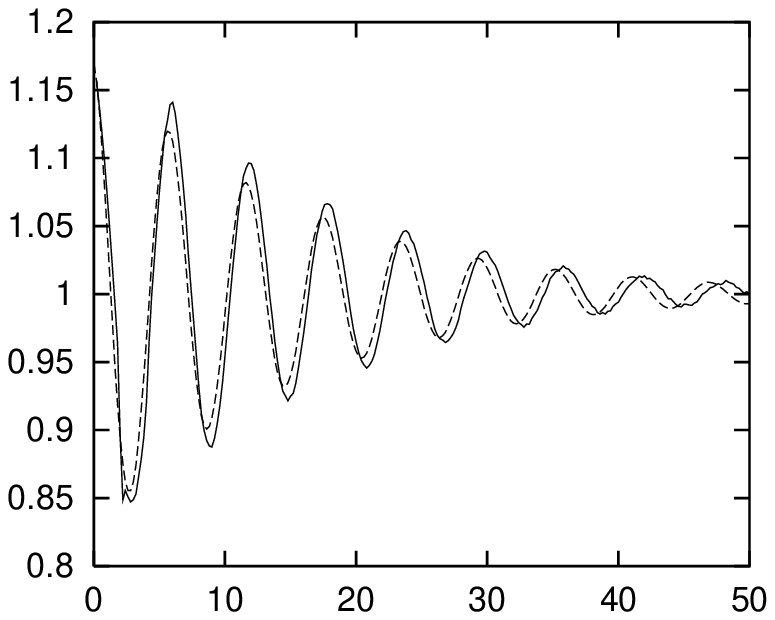}}
\put(1,4){$C(t)$}
}
\put(0,6.5){C}\put(-1,-1.5){\includegraphics{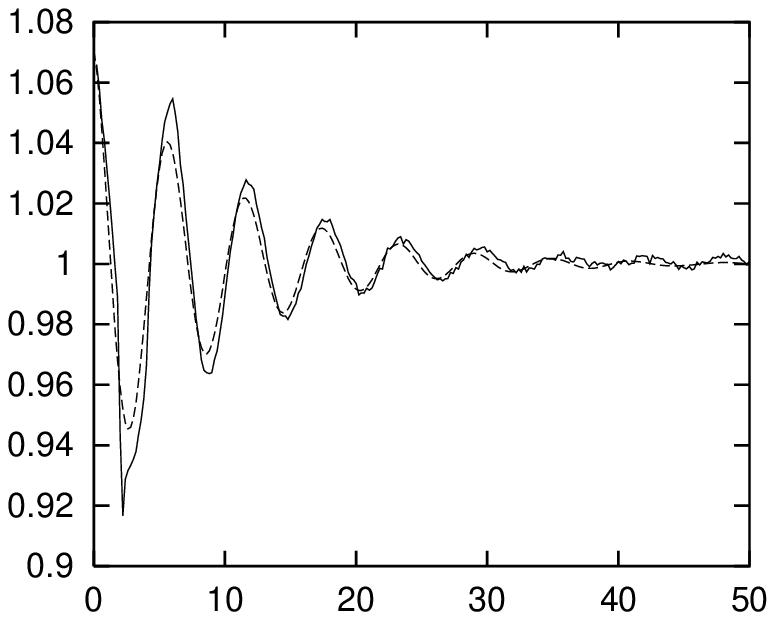}}
\put(6,0){$t$ (ms)}
\put(1,4){$C(t)$}
\end{picture}
\caption[]{AC for: A. $\sigma_{ext}=2$mV. B. $\sigma_{ext}=3$mV.
C. $\sigma_{ext}=4$mV.
Parameters as in Fig.~\ref{siparams}. Full lines: network simulation.
Dashed lines: theory (simulation of the reduced equation).}
\label{ccs}
\end{figure}

In Fig.~\ref{ccs} we compare the full AC functions
from theory (simulation of the reduced equation) 
and network simulations in three regimes,
to show the good agreement between both.

\section{Extensions}
In the previous sections a very simple network has been analyzed and
the question of the effect of some of our simplifying assumptions
legitimately arises.  In particular, we have chosen exactly identical
neurons. It can be wondered how the results are modified when some
variations in neuron properties are taken into account. In order to
address this question, we show how the previous analysis can be
generalized in two cases. Since we have seen that the oscillation
frequency is tightly linked to synaptic times, the effect of a
fluctuation in synaptic times is investigated first. We then consider
the effect of a fluctuation in the number of connections per neuron
which has been found to result in a wide spectrum of neuron steady
discharge rates (Amit and Brunel, 1997b). In both cases, it is
reassuring to find that the picture obtained from the simple model
analysis remains accurate. We finally consider a model with synaptic
currents of finite duration to analyse more precisely which time scale
plays the role of our "synaptic time" in this more realistic case.

\subsection{Effect of inhomogeneous synaptic times} 
\label{section:delays}

The analysis can easily be  
extended to the case in which time constants at each synaptic
site are drawn randomly and independently from an arbitrary probability
density function (pdf) $\Pr(\delta)$ (see Appendix \ref{app:delays}). 
In the following we consider the case of a uniform pdf between
0 and $2\delta$.

\begin{figure}
\setlength{\unitlength}{1cm}
\begin{picture}(14,6)
\put(-2,-1.5){\includegraphics{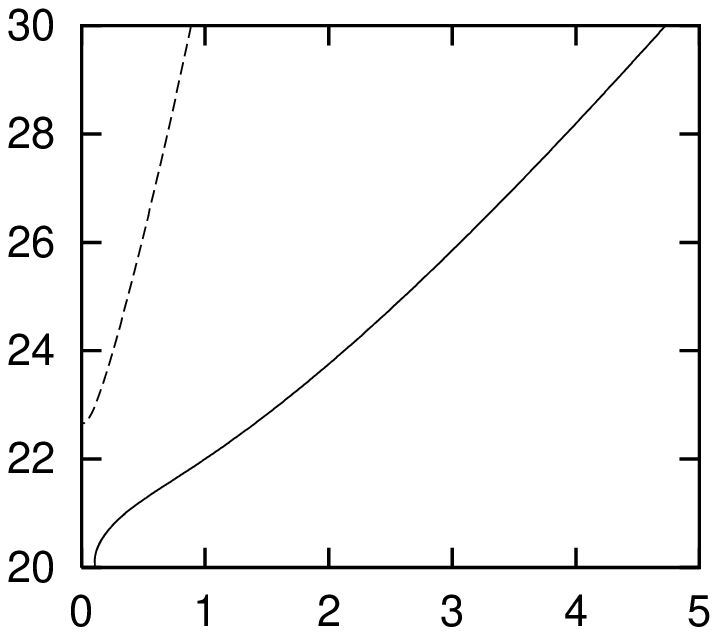}}
\put(0,3.7){$\mu _{ext}$}
\put(5.,0){$\sigma_{ext}$}
\put(2.2,5.5){OS}\put(6,3){SS}
\end{picture}
\caption{Instability line in the plane $(\mu _{ext},\sigma_{ext})$ for
$\tau=20$ms, $J=0.1$mV, $C=1000$, $\theta=20$mV, $V_r=10$mV,
$\delta=2$ms. Full line: all synaptic times equal to $\delta$.  Dashed
line: synaptic times drawn from a uniform distribution from 0 to
2$\delta$.}
\label{musidel}
\end{figure}

Fig.~\ref{musidel} shows how the instability line is modified by
random synaptic times. The region where the oscillatory instability appears
shrinks to the area above the dashed line. As the distribution of
synaptic times widens, the stationary state becomes more stable.
The introduction of random synaptic times also slightly reduces the
frequency of the oscillation.

The critical line is thus quite sensitive to the distribution
of synaptic times. In fact, distributions of synaptic times can be found such
that the stationary state is always stable (e.g.~for an exponential
distribution $\Pr(\delta) =
\exp(-\delta/\delta_0)/\delta$).

\subsection{Effect of inhomogeneous connectivity}

\begin{figure}
\setlength{\unitlength}{1cm}
\begin{picture}(14,6)
\put(-2,-1.5){\includegraphics{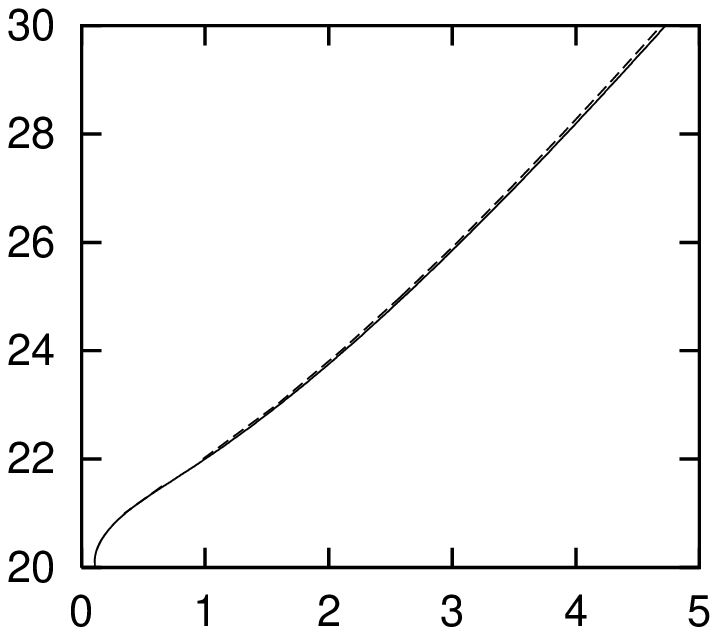}}
\put(0,3.7){$\mu _{ext}$}
\put(5.,0){$\sigma_{ext}$}
\put(3.5,5){OS}\put(6.5,3){SS}
\end{picture}
\caption{Effect of inhomogeneity in the connections on the
instability line in the plane $(\mu _{ext},\sigma_{ext})$ for
$\tau=20$ms, $J=-0.1$mV, $C=1000$, $\theta=20$mV,
$V_r=10$mV, $\delta=2$ms. Full line: all neurons receive $C$ connections.
Dashed line: connections are drawn randomly and independently at each
synaptic site with probability $C/N$.}
\label{musiinhom}
\end{figure}

\begin{figure}
\setlength{\unitlength}{1cm}
\begin{picture}(14,5.5)
\put(7.5,0){\put(-2,-1.5){\includegraphics{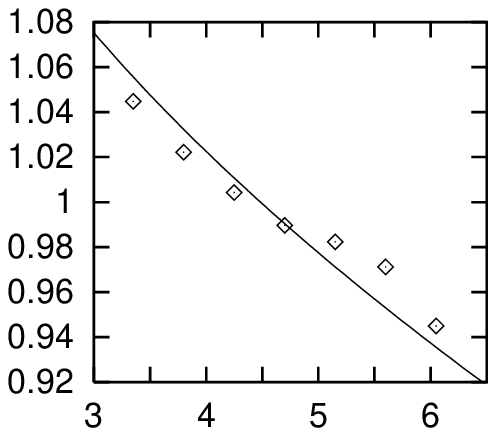}}
\put(0.,3.2){$C(\nu)$}
\put(4.,0){$\nu$ (Hz)}}
\put(0,0){\put(-2,-1.5){\includegraphics{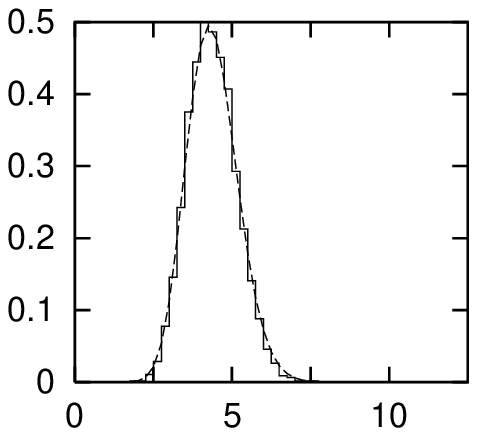}}
\put(0.2,3.2){$\Pr(\nu)$}
\put(4.,0){$\nu$ (Hz)}}
\end{picture}
\caption{Left: Distribution of spike rates (Histogram: simulation.
Dashed line: theory). The distribution is similar
to a Gaussian, unlike the distributions observed in (Amit and Brunel 1997b),
which are much wider, due to the balance between excitation and inhibition. 
Right: Relative amplitude of CC between
individual neurons and the global activity vs neuronal firing rate
(Diamonds: simulation. Full line: theory).
$\tau=20$ms, $J=-0.1$mV, $C=1000$, $\theta=20$mV,
$V_r=10$mV, $\delta=2$ms, $\mu _{ext}=25$mV, $\sigma_{ext}=2.58$mV. }
\label{ccvsrate}
\end{figure}

The analysis can also be extended to the case when the number
of connections impinging on a neuron is no longer fixed at $C$,
but rather connections are drawn at random independently at each site.
In that case the number of connections received by a neuron is 
a random variable with mean $C$ and standart deviation $\sim\sqrt{C}$.
This inhomogeneity in the connectivity provokes a significant
inhomogeneity in the individual spike rates even for $C$ large, 
because differences between the average input received by two neurons
are of the same order as the SD of the synaptic input.
The distribution of frequencies for an arbitrary network of
excitatory and inhibitory neurons has been obtained in 
(Amit and Brunel 1997b). The main steps leading to this distribution
are described in appendix \ref{app:inhomogeneous}. Next
we study how inhomogeneity affects the dynamical properties of
the network. Fig.~\ref{musiinhom} shows that the instability
line is almost unaffected by the inhomogeneity. The frequency of
the global oscillation is also very close to the one of the 
homogeneous case.

Amit and Brunel (1997b) had shown by simulations that 
the degree of synchronization of a neuron with the global activity
is strongly affected by its spike rate: neurons with low firing
frequencies tend to be more synchronized with the global activity
than neurons with high frequencies.
In appendix \ref{app:inhomogeneous} we calculate analytically the degree
of synchronization of individual neurons as a function of their
frequency. The result is shown in Fig.~\ref{ccvsrate} in which
the relative amplitude $C(\nu)$
of the cross-correlation between neurons firing at frequency $\nu$
and the global activity obtained analytically is compared with the
result of simulations. It shows indeed that low-rate neurons
are more synchronized with the global activity than high-rate neurons.
The relative amplitude of the cross-correlation between two neurons
of frequency $\nu_1$ and $\nu_2$ is  given by the product of
the two amplitudes, $C(\nu_1)C(\nu_2)$.
Note that the heterogeneity in rates and cross-correlations is not very
pronounced here, because near the critical line the fluctuations in the
external input dominate the local fluctuations, which tends to suppress
this heterogeneity.
In a network with both excitatory and inhibitory neurons with
an external excitatory input of the same order than the internal
excitatory contribution, this heterogeneity is much more pronounced
(Amit and Brunel 1997b).

\subsection{Effect of more realistic synaptic responses}

\label{section:synaptic}
Our analysis has been carried out for synaptic currents which
are described by a delta pulse. One may wonder how the
analysis generalizes for more realistic postsynaptic currents.
We consider a function $f(t)$ describing the shape of the
postsynaptic current when a spike is emitted at time $t=0$
(see e.g.~Gerstner 1995 for a review of different types of
synaptic responses).
$f(t)$ is chosen such as
$$
\int dt f(t) = 1
$$
An example often used in modelling studies and shown in
Fig.~\ref{IPSP} is the $\alpha$-function with a latency
$\tau_L$ and a characteristic synaptic time $\tau_S$:
\begin{equation}
f(t)= \left\{ \begin{array}{ll}
\frac{t-\tau_L}{\tau_S^2}\exp\left(-\frac{t-\tau_L}{\tau_S}\right)
&\mbox{for $t>\tau_L$} \\
0 & \mbox{otherwise.}
\end{array}
\right.
\end{equation}
The total synaptic current arriving at neuron $i$ is now
$$
R I_i(t) = \tau \sum_j J_{ij} \sum_k f\left(t-t_j^k\right)
$$
In the diffusion approximation the synaptic current
becomes
$$
R I_i(t) = \mu (t) + \Xi_i(t)
$$
in which the average part is given as a function of the frequency
$\nu$ and the synaptic response function $f$
by
$$
\mu (t) = \mu _{ext} -CJ\int dt' \nu(t')f(t-t') \tau.
$$
On the other hand, the fluctuating part $\Xi_i(t)$ can no longer be
approximated by a pure white noise and exhibits temporal correlations
at the scale of the width of the PSC function $f(t)$.  These temporal
correlations in the currents complicate significantly the analysis,
since the evolution of the distribution of the membrane potentials is
no longer given by a simple one-dimensional Fokker-Planck
equation. For the case of the $\alpha$-function, we would need to
solve the problem described by a three dimensional Fokker-Planck
equation. Such an analysis is beyond the scope of the present
paper. Here, we choose to ignore, as a first approximation, these
temporal correlations.  Thus we consider only the effect of the PSC
function on the average synaptic currents.  In this approximation, the
effect of the PSC function becomes equivalent to that of a
distribution of synaptic times in the delta pulse PSC case considered
in section \ref{section:delays}. For example, in the limit in which
$\tau_S$ and $\tau_L$ are small compared to the integration time
constant, the equations for the bifurcation point are
\begin{eqnarray}
G & = & \sqrt{\omega}
\left[2\frac{\tau_S}{\tau} \omega \cos
        \left(\omega\frac{\tau_L}{\tau_S}
        \right) +
        \left(1-\frac{\tau_S^2}{\tau^2}\omega^2
        \right) \sin
        \left(\omega\frac{\tau_L}{\tau}
        \right)
\right] \nonumber \\
H & = & \left(1-\frac{\tau_S^2}{\tau^2}\omega^2\right)
\left[\cos\left(\omega\frac{
\tau_L}{\tau}\right) + \sin\left(\omega\frac{\tau_L}{\tau}\right)\right]
\nonumber\\
& &+2\frac{\tau_S}{\tau}\omega \left[\cos\left(\omega\frac{\tau_L}{\tau}\right) 
-\sin\left(\omega\frac{\tau_L}{\tau}\right)\right]
\label{freqalpha}
\end{eqnarray}
In the case $\tau_L=0$ (zero latency) the equations simplify to
\begin{eqnarray}
G & = & 2\sqrt{\omega}
\frac{\tau_S}{\tau} \omega  \\
H & = & 1-\frac{\tau_S^2}{\tau^2}\omega^2
+2\frac{\tau_S}{\tau}\omega
\end{eqnarray}
In the case $H=1$, the frequency of the oscillation near the
bifurcation point is equal to $1/(\pi\tau_S)$. Note that the dependence of
the frequency on $\tau_S$ in the $\alpha$ function PSC case is 
similar to the dependence on $\delta$ in the delta pulse PSC case,
Eq.~(\ref{omega:limit}).

\begin{figure}
\setlength{\unitlength}{1cm}
\begin{picture}(14,6)
\put(-1,-1.5){\includegraphics{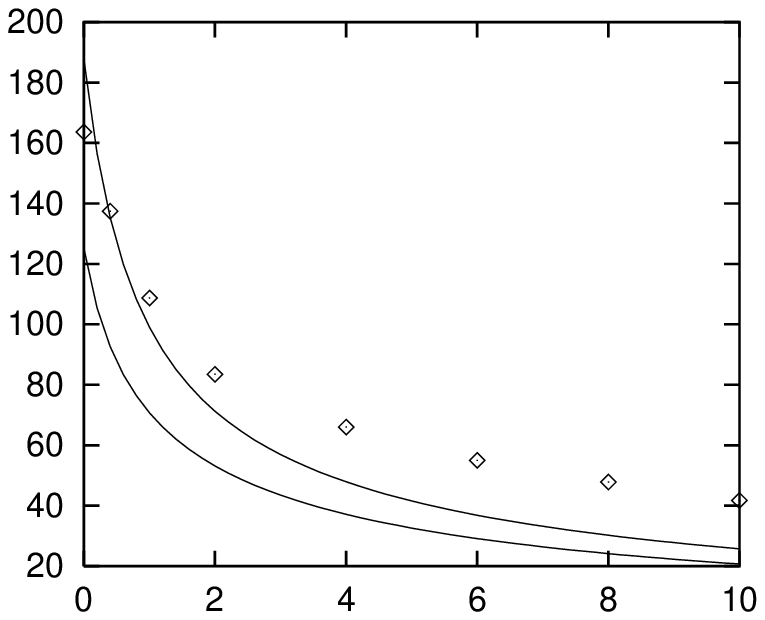}}
\put(5.5,0){$\tau_S$}
\put(1,3.5){$\frac{\omega}{2\pi}$}
\end{picture}
\caption{Dependence of the frequency of the oscillation near
the bifurcation threshold on
the synaptic decay time constant $\tau_S$, for $\tau_L=$ 2ms.
Network parameters
as in Fig.~\ref{figuresim1}. External inputs have $\mu _{ext}=$ 25mV,
$\sigma_{ext}=$ 2mV. This point is near the bifurcation line
in the whole range of $\tau_S$. $\diamond$: simulations.
Full lines: frequency given by the approximate analysis, 
Eq.~\ref{freqalpha}, for $H=1$ (lower curve),
and $H=0$ (upper curve).}
\label{omtaus}
\end{figure}

To check the validity of this approximation, we have performed
numerical simulations with fixed latency $\tau_L=$ 2ms, varying the
decay time constant of the inhibitory post synaptic currents
(IPSC) $\tau_S$. The results are shown in
Fig.~\ref{omtaus}. The approximate analysis predicts the frequency is
in the region between the two full lines (corresponding to $H=0$ and
$H=1$). Simulation results deviate from the approximate analysis
already at rather small values of $\tau_S$, because of the effect of
temporal correlations in the synaptic currents, which have the same
scale as the period of the oscillation. Nonetheless the approximation
gives a good qualitative picture of the dependence of the frequency on
$\tau_S$.

Note that the frequencies obtained in this way can be directly
compared to the data of (Whittington et al 1995, Traub et al 1996)
since the decay time constant of the PSCs can be identified with their
parameter $\tau_{GABA}$.  The frequencies obtained in the simulations
are very close to the ones obtained in that study. For example, we
obtain a frequency of about 40Hz when $\tau_S=$ 10ms, in agreement
with the in vitro recordings and the simulations of the more complex
model of (Whittington et al 1995, Traub et al 1996).  However,
one has to be careful with such a comparison, since in that {\em in
vitro} study, interneurons seem to fire at population frequency.

\section{Conclusion}

We have studied the existence of fast global oscillation in networks
where individual neurons show irregular spiking at a low rate. We have
first shown that the phenomenon can be observed in a sparsely
connected network composed of basic integrate and fire neurons. In
this very simplified setting, the phenomenon has been precisely
analyzed.  At the simplest level, it differs from other modes of
synchronisation which lead to global oscillation in that recording at
the individual neuron level shows a stochastic spike emission with
nearly Poissonian interspike intervals and little indication of the
collective behavior (see the ISI histograms in Fig.~\ref{figuresim1}).
This oscillation regime has some similarity with that obtained in Wang
et al (1995) where a hyperpolarization-activated cation current seems
to play the role of our random external inputs in generating intermittent
activity in the network.  This type of weak synchronization has
sometimes been rationalized as coming from filtering of external noise
by recurrent inhibition (Traub et al 1989 and refs.~therein). Our
analysis leads to a somewhat different picture.

We have found that, in the limit of an infinite network, the global
oscillation is due to an oscillatory instability (a supercritical Hopf
bifurcation) of the steady state.  This instability occurs at a well
defined threshold and arises from the competition between the
recurrent inhibition which favors oscillations and the intrinsic noise
in the system which tends to suppress it.
 
We have found that the global oscillation period is controlled by the
synaptic time.  This appears to agree with previous experimental
findings on slices of the rat hippocampus and with simulations results
(Whittington et al 1995, Traub et al 1996) where it is however assumed
that neurons fire at population frequency, unlike those of our
model. A similar decrease in population frequency when the GABA
characteristic time is varied is also observed in a recent {\em in
vitro} experiment in which neurons fire sparsely (Fisahn et al 1998).
More work is necessary to clarify the relative roles of the different
time constants (latency, IPSC rise time, IPSC decay time) that are
commonly used to describe the synaptic response.

The oscillation period also depends on the characteristics of the
external input, and particularly on the magnitude of the external
noise, as shown by Fig.~\ref{siparams}.  The initial rise in the
frequency when one increases $\sigma_{ext}$ followed by a saturation
at sufficiently large $\sigma_{ext}$ looks in fact similar to the
dependence of the frequency on the amount of glutamate applied to
hippocampal CA1 region {\em in vitro} (Traub et al 1996).  Our
network is in a stationary state when external inputs are
low and switches to an oscillatory regime when the magnitude of the
external inputs is increased. This phenomenon resembles
the induction of a gamma rhythm in the hippocampal slice mediated by
carbachol (Fisahn et al 1998), and  the induction of faster 200Hz
rhythms, believed to be provoked by a massive excitation of
CA1 cells through Schaeffer collaterals (see e.g.~Buzs{\'a}ki et al
1992). It is also interesting to note that a single network, with
its internal parameters fixed, is able to sustain collective
oscillations in different frequency ranges, when the characteristics
of the external input are varied.

In a finite network, the sharp transition is smoothened but the global
oscillation has different characteristics above and below the critical
threshold. Below threshold, its amplitude decreases as the network
size is increased. Above threshold, an increase in the neuron number
does not greatly modify the oscillation amplitude but increases its
coherence time.  It has been shown that the whole picture of a Hopf
bifurcation with a well-defined threshold remains accurate when some
of our simplifying assumptions are relaxed.  It would be interesting
to extend this finding to more realistic descriptions.

Our analysis also raises the important question of the synchronisation
mode used in real neural systems.  Do neocortical or hippocampal
neurons behave as oscillators with a frequency equal to the population
frequency, or irregularly with firing rates lower than the population
frequency?  In hippocampus, pyramidal cells seems clearly to be in a
irregular, low rate regime, during in vivo gamma (Bragin et al 1995),
in vivo 200Hz (Buzs{\'a}ki et al 1992) and in vitro gamma oscillations
(Fisahn et al 1998).  More recent experimental data indicates that
interneurons also typically fire at a lower frequency than the
population frequency during 200Hz oscillations in CA1 (Csicsvari
1998).
Further experimental work is needed in order to clarify
this important issue.

We have obtained a reduced description of the collective dynamics. The
analysis can certainly be extended to more complicated networks,
composed of neurons of different types or that are spatially
extended. This reduced description will hopefully prove useful in
clarifying the mechanisms of long range synchrony and in studying
propagation phenomena (Delaney et al 1996, Prechtl et al 1997).

Finally, and most importantly, the exact roles of fast oscillations
remain, at present, unclear.  Are they useful for putting in resonance
different neuronal populations as it has been suggested? Can they
serve to build a fast detector with slowly firing neurons? Are they
used as a clock mechanism? Or do they reflect the usefulness of having
a network where different neuronal populations fire in succession on a
short time scale, to code spatial information in the temporal domain?
Recent experiments (MacLeod and Laurent 1996, Stopfer et al 1997) make
us hope that elucidating the real meaning of these collective
oscillations, at least in some neural systems, is now an attainable
goal.  This is a question to which we hope to return in the future.

{\bf Acknowledgments}. We are grateful to A. Karma for discussions and
for his very stimulating role at the beginning of this work, and to
T. Bal, R. Gervais and P. Salin for informing us on real neural
networks.  N.B. is grateful to S. Fusi for useful discussions.
V.H. is glad to thank at last A. Babloyantz for an invitation to a
stimulating ESF workshop in Lanzarote which was a nice opportunity to
first learn about fast neuronal oscillations.  We thank D.~Amit and
anonymous referees for their helpful comments on the manuscript.

\appendix

\section*{Appendix}

The details of our computations are given in the following.
We have found it convenient to use the rescaled variables, 
\beq
P=\frac{2\tau\nu_0}{\sigma_0}Q,\,\,\, 
G=\frac{C J\tau\nu_0}{\sigma_0}=\frac{\mu _{0,l}}{\sigma_0},\,\,\,
H=\frac{C J^2\tau\nu_0}{\sigma_0^2}=\frac{\sigma_{0,l}^2}{\sigma_0^2},
\label{pgh}
\eeq
\beq
y=\frac{V-\mu _0}{\sigma_0},\,\,\,
y_{\theta}=\frac{\theta-\mu _0}{\sigma_0},\,\,\,
y_r=\frac{V_r-\mu _0}{\sigma_0},\,\,\,
\nu= \nu_0(1+n(t))
\label{ytrn}
\eeq
$J$ and $G$ are positive.

Using Eqs.~(\ref{pgh},\ref{ytrn}) the Fokker-Planck equation, Eq.~(\ref{fp}) becomes
\beq
\tau \frac{\partial Q}{\partial t}={\cal L}[Q] + \nu(t-\delta)
\left(G \frac{\partial Q}{\partial y} +\frac{H}{2} 
\frac{\partial^2 Q}{\partial y^2}\right)
\label{Eqfpa}
\eeq
where the linear operator $\cal L$ is defined as
$$
{\cal L}[Q] = \frac{1}{2} \frac{\partial^2 Q}{\partial y^2} + 
\frac{\partial}{\partial y}(yQ)
$$

The equation is valid on the two intervals $-\infty<y<y_r$ and 
$y_r<y<y_{\theta}$. 

The boundary conditions at
$y_r$ and $y_{\theta}$ become:
at  $y_{\theta}$
\beq
Q(y_{\theta},t)=0,\,\,\,
 \frac{\partial Q}{\partial y}(y_{\theta},t)= 
-\frac{1+n(t)}{1+H n(t-\delta)};
\label{bct}
\eeq
at $y_r$
\beq
[Q]^{y_r^+}_{y_r^-}=0,\,\,[\frac{\partial Q}{\partial y}]^{y_r^+}_{y_r^-}
=-\frac{1+n(t)}{1+H n(t-\delta)}
\label{bcr}
\eeq
(the square bracket  denotes the discontinuity of the function at $y$
namely, $[f]^{y^+}_{y^-}\equiv \lim_{\epsilon\rightarrow 0}\{f(y+\epsilon)-
f(y-\epsilon)\}$). Note the term in the r.h.s.~of Eqs.~(\ref{bct},\ref{bcr})
are identical. Thus, when we study the Fokker-Planck equation at different
orders, we will mention only the condition at $y_\theta$. The condition
at $y_r$ can be  obtained by replacing the value of the corresponding
function at $y_\theta$ by the discontinuity of the function at $y_r$.
Moreover $Q(y,t)$ should vanish sufficiently fast at $y=-\infty$ to be 
integrable.

The steady state solution obeys
\beq
{\cal L}[Q_0] = 0
\label{eqss}
\eeq
and
\beq
\frac{\partial Q_0}{\partial y}(y_{\theta})=-1,\,\,\,
[\frac{\partial Q_0}{\partial y}]^{y_r^+}_{y_r^-}
=-1
\label{bcss}
\eeq
It is given by
\beq
Q_0(y) = \left\{\begin{array}{ll}
 \exp(-y^2) \int_y^{y_{\theta}}\! du\, \exp(u^2) & y> y_r \\
 \exp(-y^2) \int_{y_r}^{y_{\theta}}\! du\, \exp(u^2) 
  & y< y_r
\end{array}\right.
\label{q0}
\eeq

From (\ref{eqss},\ref{bcss}), one easily obtains
the values of higher derivatives of $Q_0$ at $y=y_{\theta}$ and 
their discontinuities
at $y=y_{r}$, which will be used in the following, using the recurrence
relation
\beq
\frac{\partial^n Q_0}{\partial y^n}(y) = -2y 
\frac{\partial^{n-1} Q_0}{\partial y^{n-1}}(y)
-2(n-1)\frac{\partial^{n-2} Q_0}{\partial y^{n-2}}(y)
\eeq

\section{Linear stability}

\label{app:linear:stability}

The function $Q$ can be expanded around the steady state solution $Q_0(y)$
as
\beqa 
Q(y)=Q_0(y)+Q_1(y,t)+Q_2(y,t)+\cdots
\nonumber
\\
n(t)=n_1(t)+n_2(t)+\cdots
\label{dev}
\eeqa
At first order, one obtains the linear equation
\beq
\tau \frac{\partial Q_1}{\partial t}= {\cal L}[Q_1]
+ n_1(t-\delta)\left(G \frac{dQ_0}{dy} +\frac{H}{2}
\frac{d^2 Q_0}{d y^2}\right)
\label{eql}
\eeq
together with the boundary conditions
\beq
Q_1(y_{\theta},t)=0, \frac{\partial Q_1}{\partial y}(y_{\theta})= -n_1(t)
+H n_1(t-\delta)
\label{bctl}
\eeq
and
\beq
[Q_1]^{y_r^+}_{y_r^-}=0,\,\,\
[\frac{\partial Q_1}{\partial y}]^{y_r^+}_{y_r^-}
=-n_1(t)+H n_1(t-\delta)
\label{bcrl}
\eeq
Eigenmodes of (\ref{eql}) have a simple exponential behaviour in time
$$Q_1(y,t)= \exp(\lambda t/\tau)\,\hat{n}_1(\lambda)\hat{Q}_1(y,\lambda),\, n_1(t)=\exp(\lambda t/\tau)\, 
\hat{n}_1(\lambda)$$
and obey an ordinary differential equation in $y$ 
\beq
\lambda\hat{Q}_1(y,\lambda) = {\cal L}[\hat{Q}_1](y,\lambda) + e^{-\lambda \delta/\tau}
\left(G \frac{dQ_0}{dy} +\frac{H}{2}
\frac{d^2 Q_0}{d y^2}\right)
\label{eqlw}
\eeq
together with the boundary conditions
$$
\hat{Q}_1(y_{\theta},t)=0, \frac{\partial\hat{Q}_1}{\partial y}(y_{\theta})= -1
+H \exp(-\lambda\delta/\tau),
$$
and similar conditions at $y_r$.

The general solution of Eq.(\ref{eqlw}) can be written as a linear
superposition of two independent solutions $\phi_{1,2}$ of the homogeneous
equation $1/2 \phi'' + y\phi' +(1-\lambda)\phi =0$ plus a particular solution
which can be obtained by differentiating Eq.~(\ref{eqss}) with respect to $y$,
\beq
\hat{Q}_1(y,\lambda) = \left\{\begin{array}{ll}
\alpha_1^+(\lambda) \phi_1(y,\lambda) +\beta_1^+(\lambda) \phi_2(y,\lambda)
+\hat{Q}_1^p(y,\lambda)  & y> y_r \\
\alpha_1^-(\lambda) \phi_1(y,\lambda) +\beta_1^-(\lambda) \phi_2(y,\lambda)
+ \hat{Q}_1^p(y,\lambda)  & y< y_r
\end{array}\right.
\label{expq1}
\eeq
with
\beq
\hat{Q}_1^p(y,\lambda) =e^{-\lambda\delta/\tau}\left(\frac{G}{1+\lambda}
 \frac{dQ_0(y)}{dy}+ \frac{H}{2(2+\lambda)}
 \frac{d^2 Q_0(y)}{dy^2}\right)
\label{q1p}
\eeq

Solutions of the homogeneous equation $1/2 \phi'' + y\phi'
+(1-\lambda)\phi =0$ can be obtained by their series expansion around
$y=0$. They are found to be a linear combination of two functions. The
first one can be chosen as 
\beq
\phi_1(y,\lambda)=1+\sum_{n=1}^{+\infty} (-1)^n \frac
{(2y)^{2n}}{(2n)!}  \prod_{k=0}^{n-1}(k+\frac{1-\lambda}{2})
\label{phi1}
\eeq
It coincides with the confluent hypergeometric function 
$M[(1-\lambda)/2,1/2,-y^2]$ (see e.g. Abramovicz and Stegun 1970)
. A second independent
solution can also be expressed in terms 
of the hypergeometric function M as
\beq
2y M\left(1-\frac{\lambda}{2},\frac{3}{2},-y^2\right)= 2y+
\sum_{n=1}^{+\infty} (-1)^{n}\frac {(2y)^{2n+1}}{(2n+1)!}
\prod_{k=1}^{n}(k-\frac{\lambda}{2})
\label{yM}
\eeq
The asymptotic behaviour of both functions can conveniently be obtained
from the following integral representations valid for Re$(\lambda)<1/2$
\beqa
\phi_1(y,\lambda)&=&\frac{1}{\Gamma(\frac{1-\lambda}{2})}
\int_0^{+\infty}\!\!dt\,e^{-t}\,\cos(2y\sqrt{t}) t^{-\frac{1+\lambda}{2}}
\nonumber\\
2y M\left(1-\frac{\lambda}{2},\frac{3}{2},-y^2\right)&=&
\frac{1}{\Gamma(1-\frac{\lambda}{2})}
\int_0^{+\infty}\!\!dt\,e^{-t}\,\sin(2y\sqrt{t}) t^{-\frac{1+\lambda}{2}}
\label{intrep}
\eeqa
(after replacing the cosine and sine in
(\ref{intrep})
by their series expansions it is easily checked
that the obtained series in powers
of $y^n$ coincide with (\ref{phi1}) and (\ref{yM})). The following
asymptotic behaviours are found for $y\rightarrow -\infty$
\beqa
\phi_1(y,\lambda)&\sim&\frac{\sqrt{\pi}}{|y|^{1-\lambda} \Gamma(\lambda/2)}
\label{asyphy1}\\
2y M\left(1-\frac{\lambda}{2},\frac{3}{2},-y^2\right)&\sim&
- \frac{\sqrt{\pi}}{|y|^{1-\lambda} \Gamma[(1+\lambda)/2]}
\eeqa
We find it convenient to choose $\phi_2(y,\omega)$ as the particular
combination of these two functions which decays exponentially 
(i.e. like $~|y|^{-\lambda}\exp(-y^2)$) at  $y=-\infty$,
\beq
\phi_2(y,\omega) =  \frac{\sqrt{\pi}}{\Gamma\left(\frac{1+\lambda}{2}\right)}
\, M\left(\frac{1-\lambda}{2},\frac{1}{2},-y^2\right)
 + 
\frac{\sqrt{\pi}}{
\Gamma\left(\frac{\lambda}{2}\right)}\, 2 y 
M\left(1-\frac{\lambda}{2},\frac{3}{2},-y^2\right)
\eeq
Thus for $\hat{Q}_1(y,t)$ to be integrable on $[-\infty,y_{\theta}]$
we need to require $\alpha_1^- = 0$ in (\ref{expq1}).

For further reference, we give the asymptotic behaviour 
for $\lambda_2=\mbox{Im}(\lambda)\rightarrow
+\infty$,
\beqa
\phi_{1}(y,\lambda_1+i\lambda_2)&\sim& \cosh[y \sqrt{\lambda_2- i \lambda_1}(1+i)] \exp(-y^2/2)
\label{asyphi1}\\
\phi_{2}(y,\lambda_1+i\lambda_2)&\sim& 
\frac{\sqrt{\pi}}{\Gamma\left(\frac{1+\lambda}{2}\right)}
\exp[y \sqrt{\lambda_2- i \lambda_1}(1+i)-y^2/2]
\label{asyphi2}
\eeqa
where the determination of the square root is fixed by requiring it to be
positive for $\lambda_1=0$.

Finally, we note that the Wronskian $\mbox{Wr}$ of $\phi_1$ and $\phi_2$ obeys
the first order equation $\mbox{Wr}'=-2y\mbox{Wr}$ and has therefore the simple expression
\beq
\label{wronskian}
\mbox{Wr}(\phi_1,\phi_2)\equiv \phi_1 \phi_2' -\phi_1'\phi_2
= \frac{2\sqrt{\pi}}{\Gamma(\lambda/2)}\exp(-y^2)
\eeq
(the prefactor being fixed by (\ref{asyphi1},\ref{asyphi2})).

The four boundary conditions (\ref{bctl},\ref{bcrl}) give a linear 
system of four equations for the four remaining unknowns 
$\alpha_1^+, \alpha_1^-,
\beta_1^+$ and $\beta_1^-$. The condition $\alpha_1^-=0$
needed to obtain an integrable $\hat{Q}_1(y,t)$ gives 
the eigenfrequencies of
the linear equation (\ref{eql}). 
To obtain the required solvability
condition and the allied solutions, we find it convenient to use first the two
boundary conditions (\ref{bctl}) to obtain $\alpha_1^+$ and
$\beta_1^+$. This gives
\beqa
\alpha_1^+&=&\frac{1}{\mbox{Wr}(y_{\theta})}\left\{
\phi_2(y_{\theta})(1-He^{-\lambda
\delta/\tau}) - W_2\left[\hat{Q}_1^p\right](y_{\theta})
\right\}
\label{eqap1}
\\
\beta_1^+&=&-\frac{1}{\mbox{Wr}(y_{\theta})}\left\{
\phi_1(y_{\theta})(1-He^{-\lambda
\delta/\tau}) - W_1\left[\hat{Q}_1^p\right](y_{\theta})
\right\}
\label{eqbp1}
\eeqa
where $\mbox{Wr}$ denotes the Wronskian of $\phi_1$ and $\phi_2$, 
Eq.~(\ref{wronskian}), and $W_j(j=1,2)$ the Wronskian of the function
in its argument and $\phi_{1,2}$
$$W_j\left[\hat{Q}\right]\equiv
\hat{Q}\phi_j'-\hat{Q}' \phi_j \mbox{ for } j=1,2.
$$
For matters of convenience we define $\tilde\phi_{1,2}$ and $\tilde W_{1,2}$
by
$$
\tilde\phi_{1,2} = \frac{\phi_{1,2}}{\mbox{Wr}},\;\;\;\; \tilde W_{1,2}\left[\hat{Q}_1^p\right]= \frac{W_{1,2}\left[\hat{Q}_1^p\right]}{\mbox{Wr}}
$$
The two boundary conditions at $y=y_{r}$ (\ref{bcrl}) give similar
equations for  $\alpha_1^+ -\alpha_1^-$ and
$\beta_1^+- \beta_1^-$ with $y_{\theta}$ replaced by $y_{r}$
\beqa
\alpha_1^-&=&\alpha_1^+ -
\tilde\phi_2(y_r)(1-He^{-\lambda
\delta/\tau}) - \left[\tilde W_2\left[\hat{Q}_1^p\right](y)\right]\rprm 
\label{eqapm1}
\\
\beta_1^-&=&\beta_1^+ +
\tilde \phi_1(y_r)(1-He^{-\lambda
\delta/\tau}) - \left[\tilde W_1\left[\hat{Q}_1^p\right](y)\right]\rprm \nonumber
\eeqa
The two expressions (\ref{eqap1},\ref{eqapm1}) together with $\alpha_1^-=0$
give the solvability
condition and the equation for the eigenfrequencies of (\ref{eql})
\beq
\left(\tilde\phi_2(y_\theta) -
\tilde\phi_2(y_r)\right)(1-He^{-\lambda
\delta/\tau}) = \tilde W_2\left[\hat{Q}_1^p\right](y_{\theta}) -  \left[\tilde W_2\left[\hat{Q}_1^p\right](y)\right]\rprm
\label{eigeneq}
\eeq

When the synaptic time $\delta$ becomes much smaller than $\tau$, the roots $\lambda$ of
this equation become large. Considering for definiteness
roots $\lambda=\lambda_1+i \lambda_2$ with $\lambda_2>0$, in the limit $|\lambda|\rightarrow +\infty,
\lambda_2\rightarrow +\infty$, one obtains from (\ref{asyphi2}) that
$\partial_y\phi_2(y_{\theta})\gg \partial_y\phi_2(y_r)$ and
$\partial_y\phi_2(y_{\theta})\sim \sqrt{\lambda_2-i \lambda_1} (1+i) \phi_2$. 
We then note that for Eq.~(\ref{eigeneq}) to have such a root,
we need $G\sim\sqrt{|\lambda|}$. Since $H<1$ by definition, we can 
neglect the terms proportional to $H$ in $\hat{Q}_1^p$ and 
finally obtain
\beq
G\frac{e^{-\lambda
\delta/\tau}}{\lambda} \sqrt{\lambda_2-i \lambda_1} (1+i) =-1 + H e^{-\lambda
\delta/\tau}
\eeq
We focus on the root with the largest real part (together with
its complex conjugate). Its real part becomes positive, $\lambda=i \lambda_2
= i\omega_c$
when
$$
1 - H e^{-\omd}+\frac{(1-i)G e^{-\omd}}{\sqrt{\omega_c}} = 0
$$
i.e. 
$$
G = \sqrt{\omega_c}\sin\left(\omega_c\delta/\tau\right)
$$
$$
H = \sin\left(\omega_c\delta/\tau\right) +\cos\left(\omega_c\delta/\tau\right).
$$

\section{Weakly non-linear analysis}

\label{app:weakly:nl}

Our aim is to determine the lowest non-linear terms which saturate
the instability which appears when one crosses the critical line
in the plane $\mu _{ext}$, $\sigma_{ext}$.
This determines the amplitude of
the collective oscillation as well as the nonlinear contribution to its 
frequency in the vicinity of $(G_c, H_c)$. We follow
the usual strategy of pushing the development (\ref{dev}) to
higher order. One finds that the
nth-order term obey inhomogeneous linear equations with forcing terms
formed by quadratic combinations of lowest-order terms. We first determine the
second-order 
terms which are forced by quadratic combination of first-order terms
and therefore oscillate at $0$ and $2 \omega_c$. At third order,
the coupling between first and second order term generate forcing terms at
$\omega_c$ and $3 \omega_c$. While there is no problem to determine the
$3 \omega_c$ contribution, the $\omega_c$ forcing is resonant and generates
secular terms. The dynamics of the first-order terms amplitude is determined
by the requirement that it cancels the unwanted secular contribution.
The computation is not specially difficult but rather long.

We substitute the developments (\ref{dev}) of $Q(y,t)$ and $n(t)$ in 
Eq.~(\ref{Eq}) anticipating that the development parameter is of order
of the square root of the differences $G-G_c$, $H-H_c$.
Departure of $G$ from $G_c$ and of $H$ from $H_c$ will therefore only affect
the third-order terms.

The first-order terms have already been obtained, 
\beqa
Q_1(y,t)&=& e^{i\omega_c t/\tau} \hat n_1 
\hat{Q}_1(y,i\omega_c) + \mathrm{c.c.} \nonumber
\\
n_1(t)&=& e^{i\omega_c t/\tau} \hat{n}_1(i\omega_c) + \mathrm{c.c.}
\label{qn1}
\eeqa
where $\hat{Q}_1$ is given by Eqs.~(\ref{expq1},\ref{q1p},\ref{eqap1},
\ref{eqbp1}).
In Eq.~(\ref{qn1}), we recall that c.c. means that complex conjugate
terms to those explicitly written have to be added. In the following,
we omit the explicit mention of the variable $\lambda$ to lighten the notation 
since functions of $\lambda$ will all be evaluated at $i\omega_c$ (except
when explicitly specified otherwise).

By differentiation of Eq.~(\ref{eql}), one can
easily obtain recursively the values of higher
derivatives of $\hat Q_1$ at $y=y_{\theta}$ and their discontinuities
at $y=y_{r}$, which will be used in the following.

\subsection{Second order}

We first determine the second-order terms.
They  obey the equation
\beq
\tau \frac{\partial Q_2}{\partial t}= {\cal L}[Q_2] + n_2(t-\delta) 
\left(G_c \frac{dQ_0}{dy} +\frac{H_c}{2}
\frac{d^2 Q_0}{d y^2}\right)
+n_1(t-\delta) \left(G_c \frac{\partial Q_1}{\partial y} +\frac{H_c}{2}
\frac{\partial^2 Q_1}{\partial y^2}\right)
\label{eql2}
\eeq
together with the boundary conditions
\beq
Q_2(y_{\theta},t)=0, \frac{\partial Q_2}{\partial y}(y_{\theta})= -n_2(t)
+H n_2(t-\delta) -H^2 n_1^2(t-\delta)+H n_1(t) n_1(t-\delta)
\label{bct2}
\eeq
and a similar condition in $y_r$.

From (\ref{qn1}),
the forcing term on the r.h.s of Eq.~(\ref{eql2}) 
contains terms at frequencies $2\omega_c$ and $0$. Therefore, we
search $Q_2(y,t)$ and $n_2(t)$ under the form
\beqa
Q_2(y,t)&=& e^{2i\omega_c t/\tau} \hat n_1^2 \hat{Q}_{2,2}(y)+
e^{-2i\omega_c t/\tau}(\hat{n}_1^{*})^2 
\hat{Q}^{*}_{2,2}(y) + \hat{Q}_{2,0}|\hat n_1|^2
\\
n_2(t)&=& e^{2i\omega_c t/\tau} \hat n_1^2 \rho_{2,2} +
 e^{-2i\omega_c t/\tau} (\hat{n}_1^{*})^2 \rho_{2,2}^\star
+|\hat n_1|^2 \rho_{2,0}
\label{qn2}
\eeqa

Substitution of (\ref{qn2}) into (\ref{eql2}) shows that $\hat{Q}_{2,2}$ obeys
the ordinary differential equation 
\beqa
(2i\omega_c-L)\hat{Q}_{2,2}(y) & = & \rho_{2,2} e^{-2i\omega_c \delta/\tau}
\left(G_c \frac{dQ_0}{dy} +\frac{H_c}{2}
\frac{d^2 Q_0}{d y^2}\right) \nonumber \\
 & +& e^{-i\omega_c \delta/\tau}
\left(G_c \frac{\partial \hat Q_1}{\partial y} +\frac{H_c}{2}
\frac{\partial^2 \hat Q_1}{\partial y^2}\right)
\label{eql2w}
\eeqa
together with the boundary conditions
$$
\hat Q_{2,2}(y_{\theta},t)=0, \frac{\partial \hat Q_{2,2}}{\partial y}(y_{\theta})= -\rho_{2,2}
+H e^{-2\omd}\rho_{2,2} -H^2 e^{-2\omd}+H e^{-\omd}
$$
and a similar condition in $y_r$.

As above, the general solution of (\ref{eql2w}) is written as
a superposition of solution of the homogeneous equation and a
particular solution
\beq
\hat{Q}_{2,2}(y) = \left\{\begin{array}{ll}
\alpha_2^+ \phi_1(y,2i\omega_c) +\beta_2^+ \phi_2(y,2i\omega_c)
+\rho_{2,2} \hat Q_{2,2}^{so}+\hat{Q}_{2,2}^{lo}
  & y> y_r \\
\alpha_2^-\phi_1(y,2i\omega_c) +\beta_2^- \phi_2(y,2i\omega_c)
+\rho_{2,2} \hat Q_{2,2}^{so}+\hat{Q}_{2,2}^{lo}
  & y< y_r
\end{array}\right.
\label{expq}
\eeq
where
$$
\hat Q_{2,2}^{so}=e^{-2i\omega_c \delta/\tau}
\left(\frac{G_c}{1+2i\omega_c} \frac{dQ_0}{dy} +\frac{H_c}{4(1+i\omega_c)}
\frac{d^2 Q_0}{d y^2}\right)
$$
$\hat{Q}_{2,2}^{lo}$ can be obtained by
differentiation of $Q_0$ and $\hat{Q}_1$ using (\ref{eqss}) and (\ref{eqlw})
and involves only terms of lower order which have already been
determined,
\beqa
&&\hat{Q}_{2,2}^{lo}(y) = 
e^{-i\omega_c\delta/\tau} 
\left(\frac{G_c}{1+i\omega_c} \frac{\partial \hat Q_1}{\partial y} +
\frac{H_c}{2(2+i\omega_c)}
\frac{\partial^2 \hat Q_1}{\partial y^2}\right) \nonumber \\
&& -
e^{-2i\omega_c\delta/\tau}
\left(\frac{G_c^2}{2(1+i\omega_c)^2} \frac{d^2Q_0}{dy^2} +
\frac{H_c G_c}{2(1+i\omega_c)(2+i\omega_c)}
\frac{d^3 Q_0}{d y^3} \right. 
\left. +\frac{H_c^2}{8(2+i\omega_c)^2} \frac{d^4 Q_0}{dy^4}
\right) \nonumber
\eeqa

The four boundary conditions for $\hat{Q}_2$ determine the four
unknowns $\alpha_2^+,\beta_2^+,\beta_2^-,\alpha_2^-$ in terms of $\rho_{2,2}$
and the previously determined functions.
We obtain $\rho_{2,2}$ with the integrability condition $\alpha_2^-=0$
$$
\frac{\left(\tilde\phi_2(y_\theta) -
\tilde\phi_2(y_r)\right)H_c e^{-\omd}(1-H_c e^{-\omd}) + \tilde W_2\left[\hat{Q}_{2,2}^{lo}\right](y_{\theta}) -  \left[\tilde W_2\left[\hat{Q}_{2,2}^{lo}\right](y)\right]\rprm}
{\left(\tilde\phi_2(y_\theta) -
\tilde\phi_2(y_r)\right)(1-H_c e^{-2\omd})
-\tilde W_2\left[\hat{Q}_{2,2}^{so}\right](y_{\theta}) +  \left[\tilde W_2\left[\hat{Q}_{2,2}^{so}\right](y)\right]\rprm}
$$
in which all functions are taken at argument $2i\omega_c$.

The component at frequency zero $\hat{Q}_{2,0}$ obeys
\beq
0  =  {\cal L} [\hat Q_{2,0}]  +  \rho_{2,0} 
\left(G_c \frac{dQ_0}{dy} +\frac{H_c}{2}\frac{d^2 Q_0}{d y^2}\right) 
 + \left[e^{-\omd}\left(G_c \frac{\partial \hat Q_1^{\star}}{\partial y} +\frac{H_c}{2}
\frac{\partial^2 \hat Q_1^{\star}}{\partial y^2}\right) +\mbox{ c.c. }\right]
\label{eqq20}
\eeq
together with the boundary conditions
$$
\hat Q_{2,0}(y_{\theta},t)=0, \frac{\partial \hat Q_{2,0}}{\partial y}(y_{\theta})= -\rho_{2,0}(1-H)
-2 H^2 \cos(\omega_c\delta/\tau)
$$
and a similar condition in $y_r$.

Its general solution can be written
\beq
\hat{Q}_{2,0}(y) = \left\{\begin{array}{ll}
\alpha_{2,0}^+ Q_0 +\beta_{2,0}^+ \exp(-y^2)
+ \rho_{2,0}\hat{Q}_{2,0}^{so}(y) + \hat{Q}_{2,0}^{lo}(y)
  & y> y_r \\
\alpha_{2,0}^- Q_0 +\beta_{2,0}^- \exp(-y^2)
+ \rho_{2,0}\hat{Q}_{2,0}^{so}(y) +
\hat{Q}_{2,0}^{lo}(y)
  & y< y_r 
\end{array}\right.
\label{expq20}
\eeq
where
$$
\hat{Q}_{2,0}^{so}(y) = \left(G_c \frac{dQ_0}{dy} +\frac{H_c}{4}\frac{d^2 Q_0}{d y^2}\right)
$$
and it is again convenient to construct the particular solution
${Q}_{2,0}^{lo}$ by
differentiation
\beqa
\hat{Q}_{2,0}^{lo}(y) & = &
\left[ e^{+i\omega_c \delta/\tau}
\left(\frac{G_c}{1-i\omega_c} \frac{\partial \hat Q_1}{\partial y} +\frac{H_c}{2(2-i\omega_c)}
\frac{\partial^2 \hat Q_1}{\partial y^2}\right) + \mbox{ c.c. }\right] \nonumber \\
& - & 
\left(\frac{G_c^2}{1+\omega_c^2} \frac{d^2Q_0}{dy^2} +
\frac{H_c G_c(2+\omega_c^2)}{(1+\omega_c^2)(4+\omega_c^2)}
\frac{d^3 Q_0}{d y^3}  +\frac{H_c^2}{4(4+\omega_c^2)} \frac{d^4 Q_0}{dy^4}
\right)
\label{q2lo}
\eeqa

In this case, 
the four boundary conditions for $\hat{Q}_{2,0}$ are not independent
and are not sufficient to determine the four
unknowns $\alpha_{2,0}^+,\alpha_{2,0}^-,\beta_{2,0}^+,\beta_{2,0}^-$
in functions of lower order terms. This comes about because some choices
of ${Q}_{2,0}$ are  equivalent to changing the normalization of
$Q_0$. One should therefore eliminate them by imposing the condition
$\int_{-\infty}^{y_{\theta}}\! dy \hat{Q}_{2,0}=0$. In this way, one
obtains,
\beq
\rho_{2,0}= 
\frac{\left[\left(-2\frac{G_c}{1+\omega_c^2}\gamma_G
+\frac{H_c}{4+\omega_c^2}\gamma_H\right) e^{y^2}\int_{-\infty}^y\!du
e^{-u^2}\right]_{y_r}^{
y_{\theta}}+\gamma_I}
{\frac{1}{2\nu_0}+
\left[\left(G_c-\frac{H_c y}{2}\right) 
e^{y^2}\int^y_{-\infty}\! du e^{-u^2}\right]_{y_r}^{y_{\theta}}
} 
\eeq
where
$$
\gamma_G(y)=G_c y+\cos(\omega_c\delta/\tau)-
\omega_c \sin(\omega_c\delta/\tau)- \frac{H_c(2y^2+1)}{3}
$$
$$
\gamma_H(y)=4y\cos(\omega_c\delta/\tau)-
2y\omega_c \sin(\omega_c\delta/\tau)+\frac{4G_c(2y^2+1)}{3}-H_c(2y^3+3y)
$$
$$
\gamma_I= -2(y_\theta - y_r)G_c H_c \frac{2+\omega_c^2}{(1+\omega_c^2)
(4+\omega_c^2)} +\frac{H_c^2(y_\theta^2 - y_r^2)}{4+\omega_c^2}
$$
(the notation $[f]_{y_r}^{y_{\theta}}\equiv f(y_{\theta})-f(y_r)$ is used).
The derivatives of higher order of $\hat Q_{2,2}$ and $\hat Q_{2,0}$,
which are used in the following, can be obtained recursively by differentiation
of Eq.~(\ref{eql2w}) and (\ref{eqq20}).

\subsection{Third order}

We can now proceed and study the third order terms. They obey the equation
\beqa
\tau \frac{\partial Q_3}{\partial t} & = & {\cal L}[ Q_3]
+ n_3(t-\delta) \left(G_c \frac{dQ_0}{dy} +\frac{H_c}{2}
\frac{d^2 Q_0}{d y^2}\right)
\nonumber \\
&&+ n_2(t-\delta)\left(G_c \frac{\partial Q_1}{\partial y} +\frac{H_c}{2}
\frac{\partial^2 Q_1}{\partial y^2}\right)
+ n_1(t-\delta)\left(G_c \frac{\partial Q_2}{\partial y} +\frac{H_c}{2}
\frac{\partial^2 Q_2}{\partial y^2}\right) \nonumber \\
&&+ n_1(t-\delta)\left((G-G_c) \frac{dQ_0}{dy} +\frac{(H-H_c)}{2}
\frac{d^2 Q_0}{d y^2}\right)
\nonumber \\
&&
-\left\{\tau \frac{d\hat{n_1}}{dt} \hat{Q}_1 e^{i\omega_c t/\tau}
+\delta \frac{d\hat{n_1}}{dt} e^{i\omega_c (t-\delta)/\tau}
\left(G_c \frac{dQ_0}{dy} +\frac{H_c}{2}\frac{d^2 Q_0}{d y^2}\right)
+\mathrm{c.c.}\right\}
\label{eql3}
\eeqa
together with boundary conditions
\beqa
\hat Q_3(y_\theta) & = & 0 \nonumber \\
\frac{\partial \hat Q_3}{\partial y}(y_\theta) & = & -n_3(t) +H n_3(t-\delta) -2H^2 n_1(t-\delta) n_2(t-\delta) \nonumber \\
& & + H\left(n_1(t) n_2(t-\delta)+n_1(t-\delta) n_2(t)\right)+H^3 n_1^3(t-\delta) -H^2 n_1(t) n_1^2(t-\delta) \nonumber \\
& & +(H-H_c)  n_1(t-\delta) - H\delta \frac{d\hat{n_1}}{dt} 
e^{i\omega_c (t-\delta)/\tau}
\label{bc13}
\eeqa
and a similar condition holds at $y_r$.

The last two terms between brackets on the r.h.s. of (\ref{eql3})
come from the anticipation that it will be needed
to have
$\hat{n}_1$ change on a 
slow time scale to cancel secular terms. The first term arises from the
explicit time differentiation in Eq. (\ref{Eqfpa}) and does not need 
special explanations. The second is less usual and comes from the delayed
forcing $\nu(t-\delta)$ in (\ref{Eqfpa}). Formally
introducing a slow time scale $T=\epsilon t$, the delayed forcing is written
$\nu(t-\delta,T-\epsilon\delta)$.
The second term between brackets in (\ref{eql3}) is produced by
the expansion to first-order in $\epsilon$
$\nu(t-\delta,T-\epsilon\delta)=\nu(t-\delta)-\epsilon \partial_T\nu(t-\delta)
+\cdots$. The last term in the boundary condition (\ref{bc13}) appears
in the same
way.

 The forcing terms on the r.h.s. of
(\ref{eql3}) oscillate at frequencies $3\omega_c$ and $\omega_c$. Therefore,
we search $Q_3(y,t)$ and $n_3(t)$ under the form
\beqa
Q_3(y,t)&=& e^{3i\omega_c t/\tau} \hat{Q}_{3,3}(y)+e^{i\omega_c t/\tau}
\hat{Q}_{3,1}(y) + \mathrm{c.c.}
\nonumber
\\
n_3(t)&=& e^{3i\omega_c t/\tau} \hat{n}_{3,3} +
 e^{i\omega_c t/\tau} \hat{n}_{3,1}+\mathrm{c.c.}
\label{qn3}
\eeqa
We focus on the terms at frequency $\omega_c$ which are resonant with the
first order terms. They obey the equation
\beqa
&&(i\omega_c-L)\hat{Q}_{3,1}(y) =
\hat{n}_{3,1}  e^{-i\omega_c \delta/\tau}
\left(G_c \frac{dQ_0}{dy} +\frac{H_c}{2}\frac{d^2 Q_0}{d y^2}\right)
\nonumber  \\
&& + |\hat n_1|^2 \hat n_1 \left\{
\rho_{22} e^{-2i\omega_c \delta/\tau}
\left(G_c \frac{\partial \hat Q_1^{\star}}{\partial y} +\frac{H_c}{2}
\frac{\partial^2 \hat Q_1^{\star}}{\partial y^2}\right)
+\rho_{20}
\left(G_c \frac{\partial \hat Q_1}{\partial y} +\frac{H_c}{2}
\frac{\partial^2 \hat Q_1}{\partial y^2}\right) \right.
\nonumber  \\
&&
+\left.e^{-i\omega_c \delta/\tau} 
\left(G_c \frac{\partial \hat Q_{2,0}}{\partial y} +\frac{H_c}{2}
\frac{\partial^2 \hat Q_{2,0}}{\partial y^2}\right)
+ e^{i\omega_c \delta/\tau} 
\left(G_c \frac{\partial \hat Q_{2,2}}{\partial y} +\frac{H_c}{2}
\frac{\partial^2 \hat Q_{2,2}}{\partial y^2}\right)\right\}
\nonumber \\
&& +\hat{n}_1 e^{-i\omega_c \delta/\tau}
\left((G-G_c) \frac{dQ_0}{dy} +\frac{(H-H_c)}{2}\frac{d^2 Q_0}{d y^2}\right)
\nonumber \\
&&
-\tau \frac{d\hat{n_1}}{dt} \hat{Q}_1
-e^{-i
\omega_c \delta/\tau}\delta \frac{d\hat{n}_1}{dt} 
\left(G_c \frac{dQ_0}{dy} +\frac{H_c}{2}\frac{d^2 Q_0}{d y^2}\right)
\label{eql3w} 
\eeqa
The general solution of (\ref{eql3w})
can be written
\beq
\hat{Q}_3(y) = \left\{\begin{array}{ll}
\alpha_3^+ \phi_1(y,i\omega_c) +\beta_3^+ \phi_2(y,i\omega_c)
+ \hat{n}_{3,1}\hat Q_{1}^{p} + \hat Q_{3,1}^{lo}
  & y> y_r \\
\alpha_3^- \phi_1(y,i\omega_c) +\beta_3^- \phi_2(y,i\omega_c)+
\hat{n}_{3,1}\hat Q_1^p + \hat Q_{3,1}^{lo}
  & y< y_r
\end{array}\right.
\label{expq3}
\eeq
In the particular solution, $\hat Q_{1}^{p}$ is the function
that appears at first order, Eq.~(\ref{q1p}), and
as before, we can construct $\hat{Q}_{3,1}^{lo}$ by differentiation of
lower order terms:
\beq
\label{q31lo}
\hat{Q}_{3,1}^{lo} = \tau\frac{d\hat{n}_1}{dt} \hat Q_{3,1}^d + \hat{n}_1
\hat Q_{3,1}^l + \hat{n}_1|\hat{n}_1|^2 \hat Q_{3,1}^c
\eeq
where $\hat Q_{3,1}^d$ is obtained from $\hat Q_1$ by differentiation of $\phi_{1,2}$
and $\hat Q_1^p$
with respect to $\lambda$
\beq
\hat Q_{3,1}^d(y) = \left\{\begin{array}{ll}
\alpha_1^+ \partial_\lambda\phi_1(y,i\omega_c) +
\beta_1^+
\partial_\lambda\phi_2(y,i\omega_c)
+\partial_\lambda \hat Q_1^p(y,i\omega_c) & y> y_r \\
\beta_1^-\partial_\lambda\phi_2(y,i\omega_c)
+\partial_\lambda \hat Q_1^p(y,i\omega_c) & y< y_r,
\end{array}\right.
\label{q31d}
\eeq
\beq
\label{q31l}
\hat Q_{3,1}^l = e^{-\omd}\left(\frac{(G-G_c)}{1+i\omega_c} \frac{dQ_0}{dy} +\frac{(H-H_c)}{2(2+i\omega_c)}\frac{d^2 Q_0}{d y^2}\right),
\eeq
and
\beqa
\hat{Q}_{3,1}^{c}&=&
e^{\omd}
\left(\frac{G_c}{1-i\omega_c} \frac{\partial \hat Q_{2,2}}{\partial y} +\frac{H_c}{2(2-i\omega_c)}
\frac{\partial^2 \hat Q_{2,2}}{\partial y^2}\right)
\nonumber \\
&+&e^{-\omd}
\left(\frac{G_c}{1+i\omega_c} \frac{\partial \hat Q_{2,0}}{\partial y} +\frac{H_c}{2(2+i\omega_c)}
\frac{\partial^2 \hat Q_{2,0}}{\partial y^2}\right)
\nonumber \\
&+&\rho_{2,0}
\left(G_c \frac{\partial \hat Q_1}{\partial y} +\frac{H_c}{4}
\frac{\partial^2 \hat Q_1}{\partial y^2}\right)
+\rho_{2,2} e^{-2\omd}
\left(\frac{G_c}{1+2i\omega_c} \frac{\partial \hat Q_1^{\star}}{\partial y} +\frac{H_c}{4(1+i\omega_c)}
\frac{\partial^2 \hat Q_1^{\star}}{\partial y^2}\right)
\nonumber \\
&-&\frac{G_c}{1+\omega_c^2}
\left(G_c \frac{\partial^2 \hat Q_1}{\partial y^2} +\frac{H_c}{3}
\frac{\partial^3 \hat Q_1}{\partial y^3}\right)
-2\frac{H_c}{4+\omega_c^2}
\left(\frac{G_c}{3} \frac{\partial^3 \hat Q_1}{\partial y^3} +\frac{H_c}{8}
\frac{\partial^4 \hat Q_1}{\partial y^4}\right)
\nonumber \\
&-&e^{-2\omd}\frac{G_c}{1+i\omega_c}
\left(\frac{G_c}{2(1+i\omega_c)} \frac{\partial^2 \hat Q_1^{\star}}{\partial y^2} +\frac{H_c}{2(3+2i\omega_c)}
\frac{\partial^3 \hat Q_1^{\star}}{\partial y^3}\right)
\nonumber \\
&-&e^{-2\omd}\frac{H_c}{2(2+i\omega_c)}
\left(\frac{G_c}{3+2i\omega_c} \frac{\partial^3 \hat Q_1^{\star}}{\partial y^3} +\frac{H_c}{4(2+i\omega_c)}
\frac{\partial^4 \hat Q_1^{\star}}{\partial y^4}\right)
\nonumber \\
&-&\rho_{2,2} e^{-\omd}G_c\frac{2+i\omega_c}{(1-i\omega_c)(1+2i\omega_c)}
\left(\frac{G_c}{2+i\omega_c} \frac{d^2 Q_0}{dy^2} +\frac{H_c}{2(3+i\omega_c)}\frac{d^3 Q_0}{d y^3}\right)
\nonumber \\
&-&\rho_{2,0} e^{-\omd}G_c\frac{2+i\omega_c}{1+i\omega_c}
\left(\frac{G_c}{2+i\omega_c} \frac{d^2Q_0}{dy^2} +\frac{H_c}{2(3+i\omega_c)}\frac{d^3 Q_0}{d y^3}\right)
\nonumber \\
&-&\rho_{2,2} e^{-\omd}H_c\frac{4+i\omega_c}{4(1+i\omega_c)(2-i\omega_c)}
\left(\frac{G_c}{3+i\omega_c} \frac{dQ^3_0}{dy^3} +\frac{H_c}{2(4+i\omega_c)}\frac{d^4 Q_0}{d y^4}\right)
\nonumber \\
&-&\rho_{2,0} e^{-\omd}H_c\frac{4+i\omega_c}{4(2+i\omega_c)}
\left(\frac{G_c}{3+i\omega_c} \frac{d^3Q_0}{dy^3} +\frac{H_c}{2(4+i\omega_c)}\frac{d^4 Q_0}{d y^4}\right)
\nonumber \\
&+&e^{-\omd}G_c^2\frac{3+i\omega_c}{2(1-i\omega_c)(1+i\omega_c)^2}
\left(\frac{G_c}{3+i\omega_c} \frac{d^3Q_0}{dy^3} +\frac{H_c}{2(4+i\omega_c)}\frac{d^4 Q_0}{d y^4}\right)
\nonumber \\
&+&e^{-\omd}\frac{G_c H_c}{6}\left(\frac{2}{1+\omega_c^2}+\frac{4}{4+\omega_c^2}+\frac{3}{(1+i\omega_c)(2+i\omega_c)}\right)
\nonumber \\
&&\;\;\;\;\;\;\left(\frac{G_c}{4+i\omega_c} \frac{d^4Q_0}{dy^4} +\frac{H_c}{2(5+i\omega_c)}\frac{d^5 Q_0}{d y^5}\right)
\nonumber \\
&+&e^{-\omd}H_c^2\frac{6+i\omega_c}{8(2+i\omega_c)^2(2-i\omega_c)}
\left(\frac{G_c}{5+i\omega_c} \frac{d^5Q_0}{dy^5} +\frac{H_c}{2(6+i\omega_c)}\frac{d^6 Q_0}{d y^6}\right)
\label{q31c}
\eeqa

Now, upon replacing $\alpha_3^-=0$ one can try to
determine 
$\alpha_3^+,\beta_3^+,\beta_3^-,\hat{n}_3$ from the four boundary
conditions on $\hat{Q}_{3,1}(y)$. 
This provides a linear inhomogeneous system 
for the four unknowns. The inhomogeneous terms are made from
$\hat{Q}_{3,1}^{lo}(y)$ and its derivatives evaluated at $y_{\theta}$ and
$y_r$. But there is a difficulty : since we are considering the resonant
part of the third order terms, the linear operator coincides
with the $4\times 4$ matrix obtained at first order 
which has been  required to have a zero determinant. So, the equations
for $\alpha_3^+,\beta_3^+,\beta_3^-,\hat{n}_3$ are solvable only if the
inhomogeneous terms obey a solvability condition. In order to obtain
it, we find it convenient to proceed as we did at linear order
(see Eq.~(\ref{eqap1}, \ref{eqapm1})). We
obtain $\alpha_3^+$ and $\beta_3^+$ in terms of $\hat{n}_{3,1}$ and
$\hat{Q}_{3}^{lo}$ from
the $2\times 2$ system given by
the two boundary conditions at $y_{\theta}$. We then obtain similar expressions
for $\alpha_3^+$ and $\beta_3^+-\beta_3^-$. Comparing the two obtained
expressions for $\alpha_3^+$ and requiring them to be identical provides
the solvability condition,
\beq
\left(\tilde\phi_2(y_\theta) -
\tilde\phi_2(y_r)\right) \Omega = \tilde W_2\left[\hat{Q}_{3,1}^{lo}\right](y_{\theta}) -  \left[\tilde W_2\left[\hat{Q}_{3,1}^{lo}\right](y)\right]\rprm
\label{solvcond}
\eeq
where 
\beqa
\Omega & = & -\frac{\delta}{\tau}He^{-\omd}\frac{d\hat n_1}{dt}
-\hat n_1 \Omega_1 - \hat n_1|\hat  n_1|^2 \Omega_3
\nonumber\\
\Omega_1 & = & (H-H_c) e^{-\omd}
\nonumber \\
\Omega_3 & = & -2H_c^2 e^{-\omd}(\rho_{22}+\rho_{20}) +3 H_c^3 e^{-\omd}
\nonumber\\
&+&  H_c\left[\rho_{22}(e^{-2\omd}
+e^{\omd})
+\rho_{20}(1+e^{-\omd})\right] -H_c^2(2+e^{-2\omd})
\eeqa
With the help of Eqs.~(\ref{q31lo},\ref{q31d},\ref{q31l},\ref{q31c}), 
this gives the searched for equation of motion
for $\hat{n}_1$ 
\beq
\label{eqmotion}
\tau\frac{d\hat n_1}{dT} = A \hat\nu_1 - B |\hat\nu_1|^2 \hat\nu_1
\eeq
in which 
\beq
\label{a}
A = \frac{-\tilde W_2\left[\hat{Q}_{3,1}^{l}\right](y_{\theta}) +\left[\tilde W_2\left[\hat{Q}_{3,l}^{l}\right](y)\right]\rprm
-\left(\tilde\phi_2(y_\theta) -
\tilde\phi_2(y_r)\right)\Omega_1}{\tilde W_2\left[\hat{Q}_{3,1}^{d}\right](y_{\theta}) 
-\left[\tilde W_2\left[\hat{Q}_{3,1}^{d}\right](y)\right]\rprm +\left(\tilde\phi_2(y_\theta) -
\tilde\phi_2(y_r)\right)\frac{\delta}{\tau}He^{-\omd}}
\eeq
\beq
B = \frac{\tilde W_2\left[\hat{Q}_{3,1}^{c}\right](y_{\theta}) -\left[\tilde W_2\left[\hat{Q}_{3,1}^{c}\right](y)\right]\rprm
+\left(\tilde\phi_2(y_\theta) -
\tilde\phi_2(y_r)\right)\Omega_3}{\tilde W_2\left[\hat{Q}_{3,1}^{d}\right](y_{\theta}) 
-\left[\tilde W_2\left[\hat{Q}_{3,1}^{d}\right](y)\right]\rprm+\left(\tilde\phi_2(y_\theta) -
\tilde\phi_2(y_r)\right)\frac{\delta}{\tau}He^{-\omd}}
\label{b}
\eeq
These expressions simplifies in
the limit $\delta/\tau\rightarrow 0$. In the particular case
$H=0$, one obtains Eq.~(\ref{ABsimp}) of the main text.

\section{Effect of noise due to finite-size effects}

\label{app:noise}

Inserting the noise in Eq.~(\ref{eql3}),
we obtain
\beq
\label{noisymotion}
\tau\frac{d\hat n_1}{dt} = A \hat n_1 - B |\hat n_1|^2 \hat n_1 + D \sqrt{\tau} \zeta(t)
\eeq
in which $A$ and $B$ are given by Eqs.~(\ref{a},\ref{b}),
while $D$ is
\beq
\label{c}
D = \eta \frac{-\tilde W_2\left[\hat{Q}_{noise}\right](y_{\theta}) +\left[\tilde W_2\left[\hat{Q}_{noise}\right](y)\right]\rprm}
{\tilde W_2\left[\hat{Q}_{3,1}^{d}\right](y_{\theta}) 
-\left[\tilde W_2\left[\hat{Q}_{3,1}^{d}\right](y)\right]\rprm}
\eeq
where
$$
\hat{Q}_{noise} = \frac{e^{-\omd}}{1+i\omega_c} \frac{dQ_0}{dy}
$$
$\eta$ is given by Eq.~(\ref{eta}), and $\zeta$ is a complex white noise
such that
$<\zeta(t)\zeta^{\star}(t')>=\delta(t-t')$

The autocorrelation at zero time $C(0)$ is given by
$$
C(0) = 1+ 2< |\hat n_1(t)|^2 >
$$
We deduce from Eq.~(\ref{noisymotion})
the Fokker-Planck equation describing the evolution
of the p.d.f. of both real and imaginary parts of $\hat n_1$.
This equation can be converted in an equation giving
the stationary distribution $\Pr(\rho)$ of $\rho\equiv|\hat n_1|^2$. 
It satisfies
$$
\frac{\partial}{\partial \rho} \left(|D|^2 \rho
\frac{\partial \Pr}{\partial \rho}\right) = 
\frac{\partial}{\partial \rho}\left( \left[2 A_r \rho -  2 B_r
\rho^2 \right] \Pr\right)
$$
whose solution is
$$
\Pr(\rho) = \frac{\exp\left( 2 \frac{A_r}{|D|^2} \rho - 
\frac{B_r}{|D|^2} \rho^2\right)}{\int_0^{\infty} 
\exp\left( 2 \frac{A_r}{|D|^2} R - 
\frac{B_r}{|D|^2}R^2\right) dR}
$$
and the autocorrelation at zero lag is
$$
C(0) = 1 + 2 \frac{\int_0^{\infty} R\exp\left( 2 \frac{A_r}{|D|^2} R - 
\frac{B_r}{|D|^2} R^2\right) d R}{\int_0^{\infty} 
\exp\left( 2 \frac{A_r}{|D|^2} R - 
\frac{B_r}{|D|^2}R^2\right) dR}
$$
From this exact expression, it is not difficult to obtain the expressions
(\ref{c01},\ref{c02},\ref{c03}) of the main text. 

From Eq.~(\ref{noisymotion}), one can compute the behavior of the 
autocorrelation function $C(s)$. Far below the critical line, $|\hat n_1|$ is
small and the nonlinear term can be neglected. It is then easy to obtain
Eq.~(\ref{ACbelow}) of the main text.

In the oscillatory regime far above the critical line, finite size effects
provoke fluctuations
of activity around the oscillation described by Eq.~(\ref{noiselessnu1text}).
We consider a small perturbation, both in amplitude and in phase,
of the `pure' oscillation $\hat n_1\rightarrow \hat n_1 (1+r)\exp(i\phi)
$. $r$ is the perturbation in amplitude, while $\phi$ is the perturbation
in phase. To obtain the evolution equations for $r$ and $\phi$
we apply standard stochastic calculus techniques (see e.g.~Gardiner 1983,
chapter 4), and obtain,
\beqa
\label{eqrho}
\tau\dot{r} & = & -A_r  (2r +3r^2 +r^3) +\epsilon \zeta_r
+\epsilon^2 \frac{1}{2(1+r)}, \\
\label{eqphi}
\tau\dot{\phi} & = & -\frac{B_i A_r}{B_r} (2r +r^2) 
+\epsilon\frac{\zeta_i}{1+r} %-\epsilon^2 \frac{\zeta_i \zeta_r}{(1+r)^2}
\eeqa
in which $\epsilon=|D|/R$, and $\zeta_r$, $\zeta_i$ are uncorrelated
white noises. Note that the last term in the r.h.s.~of Eq.~(\ref{eqrho})
appears 
due to the fact that, upon discretizing
Eq.~(\ref{noisymotion}) with a small time step $dt$, $\phi(t+dt)-\phi(t)$
is of order $\sqrt{dt}$, not $dt$.
The calculation of the autocorrelation in terms of $r$ and $\phi$
gives, keeping only the dominant term,
$$
C(s) = 1 + 2R^2 < \cos\left((\omega_c+\Delta \omega) s/\tau + \phi(t+s)-\phi(t)\right)>.
$$
In order to calculate the autocorrelation we need to calculate
the distribution of $\Delta\phi(s)= \phi(t+s) -\phi(t)$.
From Eqs.~(\ref{eqrho},\ref{eqphi}) 
we find that, to leading order in $\epsilon$, 
it has a Gaussian distribution with mean
0 and variance
$$
\gamma^2(s) = \frac{|D|^2}{2R^2} \left[ \frac{s}{\tau} +\frac{B_i^2}{2B_r^2 A_r}
\left\{ \exp\left(-\frac{2A_r s}{\tau}\right) -1 +\frac{2A_r s}{\tau}
\right\}\right]
$$
Averaging $\cos((\omega_c+\Delta \omega) s/\tau + \Delta \phi(s))$ with such a 
distribution yields
$$
C(s) = 1 + 2R^2 \cos\left((\omega_c+\Delta \omega) s/\tau\right)\exp\left(-\gamma^2(s)/2
\right)
$$
We find a damped cosine function as
below the critical lines, but now the damping factor is no
longer a simple exponential.
For small times $s\ll \tau/(B_r R^2)$, the damping is described by 
$$
\exp\left(-\frac{\gamma^2(s)}{2}\right) \sim 
\exp\left(-\frac{|D|^2}{4R^2}\frac{s}{\tau}
\right)
$$
while for long times $s\gg \tau/(B_r R^2)$
$$
\exp\left(-\frac{\gamma^2(s)}{2}\right) \sim \exp\left(-\frac{|D|^2}{4R^2}\left(1+\frac{B_i^2}{B_r^2}\right)\frac{s}{\tau}
\right)
$$
The damping time constant in both regimes is proportional to $1/|D|^2\sim
N/C$, i.e.~to the inverse of the connection probability. When $N$ goes to
infinity at $C$ fixed the `coherence time' of the oscillation increases
linearly with $N$.

The next order in $\epsilon$ brings (after a rather tedious calculation)
a small additional contribution
to the variance, so that for long times 
$$
\exp\left(-\frac{\gamma^2(s)}{2}\right) = \exp\left(-\frac{|D|^2}{4R^2}\left(1+\frac{B_i^2}{B_r^2}\right)\frac{s}{\tau}\left[1
+\frac{|D|^2}{2 A_r} + O\left(|D|^4\right)\right]
\right)
$$

\section{Randomly distributed synaptic times}

\label{app:delays}
The calculations performed in the case in which all synaptic times have the same
value
can be repeated in the more general situation in which synaptic times are drawn 
randomly and independently
at each site with distribution $\Pr(\delta)$.
The difference is that, in all equations were functions of $\delta$ appears,
we need to integrate these functions with the p.d.f.$\Pr(\delta)$.
For example, we find that the critical line where the instability appears
is given by
\beq
\left(\tilde\phi_2(y_\theta) -
\tilde\phi_2(y_r)\right)\left(1-H\int\Pr(\delta)e^{-w
\delta/\tau}d\delta\right) = \tilde W_2\left[\hat{Q}_1^p\right](y_{\theta}) -  \left[\tilde W_2\left[\hat{Q}_1^p\right](y)\right]\rprm
\label{eigeneq:inhomdelay}
\eeq
in which
\beq
\hat{Q}_1^p(y,w) =\int\Pr(\delta) e^{-w\delta/\tau}d\delta\left(\frac{G}{1+w}
 \frac{dQ_0(y)}{dy}+ \frac{H}{2(2+w)}
 \frac{d^2 Q_0(y)}{dy^2}\right)
\eeq

\section{Inhomogeneous networks}

\label{app:inhomogeneous}
We now relax the constraint that the number of connections received
by a neuron be precisely equal to $C$. The connections are randomly
and independently drawn at each possible site. They are present
with probability $C/N$.
In this situation, the dynamics of different neurons 
will depend on
this number of connections they receive: this number is now a random
variable with mean $C$ and variance $C(1-\epsilon)$. For example,
their frequency will be a decreasing function of the number
of connections.
The connectivity matrix
is defined by $J_{ij}=J e_{ij}$ where for all $i,j$ $e_{ij}=1$
with probability $\epsilon$.
The distribution of frequencies in the stationary state in
such a situation has been obtained, for the case of a network
with both excitatory and inhibitory neurons, by (Amit and Brunel 1997b).
The distribution of stationary frequencies can be obtained
as a special case of this analysis. We briefly recall
here the main steps of this analysis, before turning to the stability
analysis.

Averaging the synaptic input only on the randomness of spike
emission times of presynaptic neurons, we get that the mean and
the variance of local inputs are given by
$$
\mu _i = J\tau \sum_{j} e_{ij} \nu_j,  \;\;\; 
\sigma_i^2=J^2 \tau\sum_{j} e_{ij} \nu_j.
$$
Since the number of inputs to each neuron is very large, the 
spatial distribution
of the variable $\sum_{j} e_{ij} \nu_j$, which determines completely
the spatial distribution of $\mu $ and $\sigma$, will be close to a
Gaussian whose two first moments can be calculated as a function
of the two first moments of the spatial distribution of frequencies:
$$
< \sum_{j} e_{ij} \nu_j > = C \overline\nu
$$
$$
< \left(\sum_{j} e_{ij} \nu_j - C\overline\nu\right)^2 > =
C\left(\overline{\nu^2}-\epsilon \overline\nu^2\right)
$$
Thus the variable
$$
z_i = \frac{\sum_{j} e_{ij} \nu_j - 
C \overline\nu}{\sqrt{C\left(\overline{\nu^2}-\epsilon \overline\nu^2\right)}}
$$
has a Gaussian distribution, $\rho(z)=\exp(-z^2/2)/\sqrt{2\pi}$.
Thus a neuron receives, with probability $\rho(z)$, a local input with
moments
\beq
\label{muz}
\mu (z)= -J\tau(C \overline\nu +z\sqrt{C\left(\overline{\nu^2}-
\epsilon \overline\nu^2\right)})
\eeq
and
\beq
\label{sigmaz}
\sigma^2(z)=J^2 \tau(C \overline\nu +z\sqrt{C\left(\overline{\nu^2}-\epsilon \overline\nu^2\right)})
\eeq

\subsection{Distribution of frequencies in stationary state}

In the stationary state the frequency of a neuron with moments
$\mu (z)$ and $\sigma(z)$ is given by
\beq
\label{nuz}
\nu_0(z)= \left(\tau\sqrt{\pi} 
\int_{\frac{V_r-\mu (z)}{\sigma(z)}}^{\frac{\theta-\mu (z)}{\sigma(z)}} du 
\exp(u^2) (1+\mbox{erf}(u))\right)^{-1}
\eeq
The two first moments of the distribution of frequencies can
then be determined in a self-consistent way, using
$$
\overline\nu_0 = \int dz \rho(z) \nu_0(z),\;\;\;
\overline{\nu_0^2} = \int dz \rho(z)\nu_0^2(z)
$$
These equations, together with Eqs.~(\ref{muz},\ref{sigmaz},\ref{nuz}), 
fully determine the whole
distribution of stationary frequencies, which can be obtained
using the relation
$$
P(\nu) = \int dz \rho(z) \delta(\nu - \nu_0(z))
$$

\subsection{Linear stability analysis}

The linear stability analysis of Section \ref{app:linear:stability} 
can be generalized to
the inhomogeneous network. We give here the main steps of this analysis.

We expand the frequencies around the stationary frequency,
$$
\nu(z) = \nu_0(z)\left(1+n_1(z,t)+\ldots\right),
$$
and, defining for each $z$ $y= (x-\mu _0(z))/\sigma_0(z)$,
$$
P = \frac{2\tau\nu_0(z)}{\sigma_0(z)}(Q_0(y,z)+Q_1(y,z,t)+\ldots)
$$
The moments of the spatial distribution of frequencies can be
expanded in the same way,
$$
\overline\nu = \overline\nu_0\left(1+\overline n_1(t)+\ldots\right),
$$
$$
\overline{\nu^2} = \overline{\nu_0^2}\left(1+\overline{n_1^2(t)}+\ldots\right),
$$
where
$$
\overline n_1(t) = \frac{1}{\overline\nu_0}\int dz \rho(z) \nu_0(z) n_1(z,t)
$$
$$
\overline{n_1^2}(t) = \frac{2}{\overline{\nu_0^2}}\int dz \rho(z) \nu_0^2(z) n_1(z,t)
$$

The Fokker-Planck equation at first order is
\beqa
\tau \frac{\partial Q_1}{\partial t} & = &{\cal L}[Q_1]
+\frac{\left(H_1(z)\overline n_1(t-\delta) +H_2(z)\overline{n_1^2}(t-\delta)\right) }{2}
\frac{\partial^2 Q_1}{\partial y^2} \nonumber \\ 
&& +\left(G_1(z)\overline n_1(t-\delta) +G_2(z)\overline{ n_1^2}(t-\delta)\right)
\frac{\partial Q_1}{\partial y} 
\label{fpinhom}
\eeqa
where
$$
G_1(z)= \frac{JC\overline\nu_0\tau -\epsilon J\tau z\sqrt{C\left(\overline{\nu_0^2}-\epsilon \overline\nu_0^2\right)}\frac{\overline{\nu_0^2}}{\overline{\nu_0^2}-\epsilon \overline\nu_0^2}}{\sigma_0(z)}
$$
$$
G_2(z) = \frac{J\tau z\sqrt{C\left(\overline{\nu_0^2}-\epsilon \overline\nu_0^2\right)}\frac{\overline{\nu_0^2}}{\overline{\nu_0^2}-\epsilon \overline\nu_0^2}}{2\sigma_0(z)}
$$
$$
H_1(z)= \frac{J^2C\overline\nu_0\tau -\epsilon J^2\tau z\sqrt{C\left(\overline{\nu_0^2}-\epsilon \overline\nu_0^2\right)}\frac{\overline{\nu_0^2}}{\overline{\nu_0^2}-\epsilon \overline\nu_0^2}}{\sigma_0^2(z)}
$$
$$
H_2(z) = \frac{J^2\tau z\sqrt{C\left(\overline{\nu_0^2}-\epsilon \overline\nu_0^2\right)}\frac{\overline{\nu_0^2}}{\overline{\nu_0^2}-\epsilon \overline\nu_0^2}}{2\sigma_0^2(z)}
$$

The eigenmodes of Eq.~(\ref{fpinhom}) can be written
$$
Q_1(y,z,t)= \hat Q_1(y,z) \exp(i\omega t/\tau) + \mbox{c.c.}
$$
$$
n_1(z,t) = \hat n_1(z)\exp(i\omega t/\tau) + \mbox{c.c.}
$$
leading to the solvability conditions, for each $z$
$$
\hat n_1(z) = I(z) \overline{\hat n_1} +J(z) \overline{\hat n_1^2}
$$
where
$$
I(z)= \frac{\tilde W_2\left[R_1\right](y_{\theta}) -  
\left[\tilde W_2\left[R_1\right](y)\right]\rprm+H_1(z) e^{-i\omega
\delta/\tau}\left(\tilde\phi_2(y_\theta) -
\tilde\phi_2(y_r)\right)}{\tilde\phi_2(y_\theta) -
\tilde\phi_2(y_r)}
$$
$$
J(z)= \frac{\tilde W_2\left[R_2\right](y_{\theta}) -  \left[\tilde W_2\left[R_2\right](y)\right]\rprm+H_2(z) e^{-i\omega
\delta/\tau}\left(\tilde\phi_2(y_\theta) -
\tilde\phi_2(y_r)\right)}{\tilde\phi_2(y_\theta) -
\tilde\phi_2(y_r)}
$$
with
\beq
R_{1,2} =e^{-i\omega\delta/\tau}\left(\frac{G_{1,2}(z)}{1+iw}
 \frac{dQ_0(y)}{dy}+ \frac{H_{1,2}(z)}{2(2+iw)}
 \frac{d^2 Q_0(y)}{dy^2}\right)
\eeq
Multiplying the above equation by $\rho\nu_0$ (2$\rho\nu_0^2$)
and integrating with respect to $z$ we obtain
$$
\overline{\hat n_1} = 
\frac{<\nu_0 I>}{\overline{\nu_0}} \overline{\hat n_1} +
\frac{<\nu_0 J>}{\overline{\nu_0}} \overline{\hat n_1^2}
$$
$$
\overline{\hat n_1^2} = 2\frac{<\nu_0^2 I>}{\overline{\nu_0^2}}\overline{\hat n_1} +
2\frac{<\nu_0^2 J>}{\overline{\nu_0^2}} \overline{\hat n_1^2}
$$
where we use the notation $<\ldots>= \int dz\rho(z)\ldots$.
The instability point together with the associated frequency
are given by the condition that the associated determinant
vanishes, i.e.
$$
1= \frac{<\nu_0 I>}{\overline{\nu_0}}+ 2\frac{<\nu_0^2 J>}{\overline{\nu_0^2}}
+2 \frac{<\nu_0^2 I><\nu_0 J> - <\nu_0^2 J><\nu_0 I>}{\overline{\nu_0}\overline{\nu_0^2}}
$$

The relative degree of synchrony of population $z$ with the
collective oscillation is
given by
$$
\hat n_1(z) = \overline{\hat n_1} \left(I(z) + J(z) \frac{2\frac{<\nu_0^2 I>}{\overline{\nu_0^2}}}{\left(1-2\frac{<\nu_0^2 J>}{\overline{\nu_0^2}}\right)}\right)
$$

\subsection*{References}

\begin{description}

\item[]
Abbott LF and Van Vreeswijk C 1993
Asynchronous states in a network of pulse-coupled oscillators,
{\em Phys. Rev. E} 48 1483

\item[]
Abeles M 1991 {\em Corticonics}, (New York: Cambridge University Press)
 
\item[]
Abramowitz M and Stegun IA 1970 {\em Tables of
Mathematical Functions}  (Dover Publications, NY).

\item[]
Amit DJ and Brunel N 1997a A model of global spontaneous activity
and local delay activity during delay periods in the cerebral cortex,
{\em Cerebral Cortex} 7 237

\item[]
Amit DJ and Brunel N 1997b Dynamics of recurrent networks
of spiking neurons before and after learning, {\em Network}, 8 373

\item[]
Bragin A, Jando G, Nadasdy Z, Hetke J, Wise K and Buzs{\'a}ki G
1995, Gamma (40-100 Hz) oscillation in the hippocampus of the
behaving rat, {\em J. Neurosci.} 15 47

\item[]
Braitenberg V and Sch{\"u}tz A 1991 {\em Anatomy of cortex}, 
(Springer-Verlag, Berlin).

\item[]
Bender CM and Orszag SA 1987, {\em Advanced Mathematical Methods for
Scientists and Engineers} (Mc Graw-Hill, Singapore).

\item[]
Buzs{\'a}ki G and Chrobak JJ 1995, Temporal structure in spatially
organized neuronal ensembles: a role for interneuronal networks,
{\em Current Opinion in Neurobiology} 5 504

\item[]
Buzs{\'a}ki G, Horvath Z, Urioste R, Hetke J and Wise K 1992,
High frequency network oscillation in the hippocampus, {\em Science}
256 1025

\item[]
Chandrasekhar S 1943 Stochastic Problems in Physics and Astronomy
{\em Rev. Mod. Phys.} 15 1

\item[]
Csicsvari, J., H.~Hirase, A.~Czurko, and G.~Buzs{\'a}ki (1998).
Reliability and state dependence of pyramidal cell-interneuron
synapses in the hippocampus: an ensemble approach in the behaving rat.
{\em Neuron\/} 21, 179--189.

\item[]
Delaney KR, Gelperin A, Fee MS, Flores JA, Gervais R, Tank DW
and Kleinfeld D 1994 Waves and stimulus-modulated dynamics in an oscillating
olfactory network {\em Proc. Natl. Acad. Sci. USA} 91 669-673

\item[]
Eckhorn R, Frien A, Bauer R, Woelbern T and Kehr H
1993 High frequency (60-90 Hz) oscillations in primary visual
cortex of awake monkey , {\em NeuroReport}  4 243-246

\item[]
Fisahn, A., F.~G. Pike, E.~H. Buhl, and O.~Paulsen (1998).
Cholinergic induction of network oscillations at 40hz in the
hippocampus {\em in vitro}.
{\em Nature\/} 394, 186--189.

\item[]
Gardiner CW 1983 {\em Handbook of stochastic methods}
(Spinger-Verlag, Berlin)

\item[]
Gerstner W 1995, Time structure of the activity in neural network models,
{\em Phys. Rev. E} 51 738-758

\item[]
Gerstner W, van Hemmen JL and Cowan JD 1996 What matters in
neuronal locking? {\em Neural Computation} 8 1653-1676

\item[]
Golomb D and Rinzel J 1994 Clustering in globally coupled inhibitory neurons,
{\em Physica D} 72 259-282

\item[]
Gray CM 1994 Synchronous oscillations in neuronal systems:
mechanisms and functions {\em J. Comput. Neurosci.}  1 11-38

\item[]
Gray CM, K{\"o}nig P, Engel AK and Singer W 1989 Oscillatory responses
in cat visual cortex exhibit inter-columnar synchronization which reflects
global stimulus patterns, {\em Nature} 338 334

\item[]
Gray CM and McCormick DA 1996, Chattering cells: superficial pyramidal
neurons contributing to the generation of synchronous oscillations in
the visual cortex {\em Science} 274 109

\item[]
Hansel D, Mato G and Meunier C 1995, Synchrony in excitatory 
neural networks, {\em Neural Computation} 7 307

\item[]
Hirsch MW and Smale S 1974, {\em Differential Equations, Dynamical
Systems and Linear Algebra} (Academic Press, New York)

\item[]
Kopell N and LeMasson G 1994, Rhythmogenesis, amplitude modulation,
and multiplexing in a cortical architecture, {\em
Proc. Natl. Acad. Sci. USA}, 91, 10586--10590

\item[]
Kreiter AK and Singer W 1996, Stimulus-dependent synchronization
of neuronal responses in the visual cortex of the awake macaque
monkey, {\em J. Neurosci.} 16 2381

\item[]
Laurent G and Davidowitz H 1994 Encoding of olfactory information
with oscillating neural assemblies, {\em Science} 265 1872

\item[]
MacLeod K and Laurent G 1996 Distinct mechanisms for synchronization and 
temporal patterning of odor-encoding neural assemblies,
{\em Science} 274 976-979

\item[]
Mirollo RE and Strogatz SH 1990 Synchronization of pulse-coupled
biological oscillators, {\em SIAM J. of Appl. Math.}  50 1645

\item[]
Prechtl JC, Cohen LB, Pesaran B, Mitra PP and Kleinfeld D 1997 
Visual stimuli induce waves of electrical activity
in turtle cortex {\em Proc. Natl. Acad. Sci. USA} 94 7621-7626

\item[]
Rappel WJ and Karma A 1996
Noise-Induced Coherence in Neural Networks
{\em  Phys. Rev. Lett.} 77 3256-3259

\item[] 
Ritz R and Sejnowski TJ 1997 
Synchronous oscillatory activity in sensory systems: new vistas on
mechanisms
{\em Current opinion in Neurobiology} 7 536-546

\item[]
Sakaguchi H, Shinomoto S and Kuramoto Y 1988 
Phase transitions and their bifurcation analysis in a large population
of active rotators with mean-field coupling,
{\em Prog. Theor. Phys.} 79 600-607

\item[]
Singer W and Gray CM 1995 Visual feature integration and the temporal
correlation hypothesis, {\em Ann. Rev. Neurosci.} 18 555

\item[]
Stopfer M, Bhagavan S, Smith BH and Laurent G 1997
Impaired odour discrimination on desynchronization of odour-encoding neural
assemblies, {\em Nature} 390 70-74

\item[]
Strogatz SH and Mirollo RE 1991
Stability of incoherence in a population of coupled oscillators,
{\em J.~Stat.~Phys.} 63 613-635

\item[]
Traub RD, Miles R and Wong RKS 1989 Model of the origin of rhythmic
population oscillations in the hippocampal slice,
{\em Science} 243 1319

\item[]
Traub RD, Whittington MA, Colling SB, Buzs{\'a}ki G and Jefferys JGR 1996
Analysis of gamma rhythms in the rat hippocampus {\em in vitro}
and {\em in vivo} {\em J. Physiol.} 493 471

\item[]
Treves, A. (1993).
Mean-field analysis of neuronal spike dynamics.
{\em Network\/} 4, 259-284

\item[]
Tsodyks M and Sejnowski T 1995 Rapid state switching in
balanced cortical network models, {\em Network} 6 111-124

\item[]
Van Vreeswijk C, 1996
Partial synchronization in populations of pulse-coupled oscillators,
{\em Phys. Rev. E} 54 5522-5537

\item[]
Van Vreeswijk C, Abbott L and Ermentrout GB, 1994 
When inhibition not excitation
synchronizes neural firing, {\em J. Comput. Neurosci.} 1 313

\item[]
Van Vreeswijk C and Sompolinsky H 1996 Chaos in neuronal networks with
balanced excitatory and inhibitory activity, {\em Science} 274,
1724-1726

\item[]
Wang X-J and Buzs{\'a}ki G 1996, Gamma oscillation by synaptic inhibition
in a hippocampal interneuronal network model,
{\em J. Neurosci.} 16 6402

\item[]
Wang X-J, Golomb D and Rinzel J 1995, Emergent spindle oscillations
and
intermittent burst firing in a thalamic model: Specific neuronal
mechanisms, {\em Proc. Natl. Acad. Sci. USA} 92 5577--5581

\item[]
Whittington MA, Traub RD, Jefferys JGR 1995 Synchronized oscillations
in interneuron networks driven by metabotropic glutamate receptor activation,
{\em Nature} 373 612

\item[]
Ylinen A, Bragin A, Nadasdy Z, Jando G, Szabo I, Sik A and Buzs{\'a}ki G
1995, Sharp-wave associated high frequency oscillation (200 Hz)
in the intact hippocampus: network and intracellular mechanisms,
{\em J. Neurosci.} 15 30

\end{description}

\end{document}